\documentclass[aps,amsmath,amssymb,nofootinbib,twocolumn]{revtex4}

\usepackage{graphicx}
\usepackage{pgfplots}
\pgfplotsset{compat=1.17}
\usepackage{bm}
\usepackage{ragged2e}
\usepackage{xcolor}
\usepackage{tikz}
\usetikzlibrary{arrows.meta,calc,positioning}
\usepackage{hyperref}

\definecolor{qhblue}{RGB}{0,114,178}
\definecolor{qhorange}{RGB}{230,159,0}
\definecolor{qhred}{RGB}{213,94,0}
\definecolor{qhgreen}{RGB}{0,158,115}
\definecolor{qhgray}{RGB}{105,105,105}

\tikzset{
  qhframe/.style={
    draw=black!75,
    rounded corners=2pt,
    line width=0.6pt,
    fill=white,
    inner sep=4pt
  },
  qhsite/.style={
    circle,
    draw=black!80,
    line width=0.65pt,
    fill=white,
    minimum size=11mm,
    inner sep=1.5pt,
    align=center
  },
  qhsmallsite/.style={
    circle,
    draw=black!75,
    line width=0.6pt,
    fill=white,
    minimum size=8mm,
    inner sep=1pt,
    align=center
  },
  qhmap/.style={
    -{Latex[length=2.1mm,width=1.35mm]},
    line width=0.75pt,
    draw=black!80
  },
  qhforward/.style={
    -{Latex[length=1.9mm,width=1.25mm]},
    line width=0.9pt,
    draw=qhblue
  },
  qhreverse/.style={
    -{Latex[length=1.8mm,width=1.2mm]},
    line width=0.65pt,
    draw=qhgray
  },
  qhcurrent/.style={
    -{Latex[length=2.0mm,width=1.3mm]},
    line width=1.0pt,
    draw=qhorange
  },
  qhlink/.style={
    line width=0.8pt,
    draw=black!80
  },
  qhaxis/.style={
    -{Latex[length=1.8mm,width=1.15mm]},
    line width=0.65pt,
    draw=black!85
  }
}

\begin{document}
\title{Exceptional structure and Lifshitz criticality in a non-Hermitian Kitaev chain}
\author{Aritra Ghosh$^{1}$\footnote{aritraghosh500@gmail.com}, Akash Sinha$^2$, and Miloslav Znojil$^{3,4}$}
\affiliation{$^1$Center for Theoretical Physics, Polish Academy of Sciences, Warsaw, Poland\\
$^2$School of Basic Sciences, Indian Institute of Technology Bhubaneswar, Argul, Jatni, Khurda, Odisha 752050, India\\
$^3$The Czech Academy of Sciences, Nuclear Physics Institute, \v{R}e\v{z} 292,  250 68 Husinec, Czech Republic\\
$^4$Department of Physics, Faculty of Science, University of Hradec Kr\'{a}lov\'{e}, Rokitansk\'{e}ho 62, 50003 Hradec Kr\'{a}lov\'{e}, Czech Republic}
\vskip-2.8cm
\date{\today}
\vskip-0.9cm

\vspace{5mm}
\begin{abstract}
We study an exactly solvable non-Hermitian $N$-site Kitaev chain with unequal pair-creation and pair-annihilation amplitudes and vanishing intersite hopping. We show that the chain supports an exceptional point (EP) at which all the many-body energies collapse to zero and the largest Jordan block grows quadratically with $N$. By constructing an explicit positive metric for the associated $N$-dimensional one-body problem which admits an extension to the full many-body Hamiltonian, we demonstrate that this EP collapse can be approached through a continuous quasi-Hermitian path in the parameter space. Further, in a different regime with onsite energies alternating between two unequal positive values, we show that pairing amplitudes of opposite signs can lead to crossings between ground states of opposite fermion parity within a quasi-Hermitian regime. Their accumulation in the thermodynamic limit produces a Lifshitz transition, characterized by particle-/hole-pocket formation, quadratic critical dispersion, and a nonanalytic ground-state energy density. We show that this transition remains separated from the EP boundary and ceases to exist for balanced real pairing.
\end{abstract}
\maketitle

\section{Introduction}
Non-Hermitian Hamiltonians ($H \neq H^\dagger$) have garnered a significant amount of interest in the recent decades, both from the open-system perspective \cite{Rotter_2009} and also from the study of pseudo-Hermitian, specifically, quasi-Hermitian operators \cite{Scholtz_1992,Mostafazadeh_2002,Bender_2002,Faria_2006,Znojil_2008,Znojil_2009,Mostafazadeh_2010,Bagarello_2015}. Focusing on the latter perspective, a Hamiltonian operator $H$ is said to be quasi-Hermitian if one can write
\begin{equation}
H^\dagger = \Theta H \Theta^{-1},
\label{QH_def}
\end{equation} 
where $\Theta$ is a positive, Hermitian, and invertible operator \cite{Scholtz_1992}. Restricting ourselves to finite-dimensional systems, Eq. (\ref{QH_def}) for a positive $\Theta$ guarantees a real spectrum and diagonalizability; if $\Theta$ is instead allowed to be indefinite, the reality of the spectrum is no longer guaranteed though nonreal eigenvalues must occur in complex-conjugate pairs \cite{Mostafazadeh_2002,Mostafazadeh_2010}. Keeping our attention to positive $\Theta$ and hence real eigenvalues, the corresponding time evolution is unitary when viewed in a Hilbert space where the inner product carries $\Theta$ as the metric operator, which also leads to positive norms of states due to the positivity of $\Theta$. As a matter of fact, one may choose an invertible ``Dyson'' map $\Omega$ as
\begin{equation}
h=\Omega H\Omega^{-1}=h^\dagger,
\label{Dyson_def}
\end{equation}
with $\Theta=\Omega^\dagger\Omega$. The canonical choice $\Omega=\Theta^{1/2}$ is positive and Hermitian \cite{Mostafazadeh_2010}, in which case $\Theta=\Omega^2$. While $h$ shares the same spectrum as $H$ in the domain of validity of the map in Eq. (\ref{Dyson_def}), neither $h$ nor $\Omega$ are unique \cite{Ghosh_2026}. But that poses no problem, for all such finite-dimensional Hermitian representatives are isospectral and hence unitarily equivalent. Physical equivalence, however, also requires a consistent identification of the states and observables \cite{Ghosh_2026}.

\vspace{2mm}

A striking feature of non-Hermitian systems is the existence of exceptional points (EPs) at which the Hamiltonian becomes nondiagonalizable \cite{Berry_2004,Heiss_2012}. Throughout this work, an EP of order $n$, denoted EP$n$, means that the Jordan normal form contains an $n$-dimensional Jordan block at the coalesced eigenvalue. Such non-Hermitian degeneracies have been observed experimentally \cite{Dembowski_2001,Lee_2009,Kim_2016,Bergman_2021} and have been associated with intriguing topological features \cite{Li_2018,Kunst_2018,Ghatak_2019,Ghatak_2020,Bergholtz_2021}, while the emergence of Hermitian topology from non-Hermitian knots has been described in \cite{Hajong_2026}. Since EPs are points at which a non-Hermitian operator ceases to be diagonalizable because some or all of its eigenstates coalesce \cite{Heiss_2012}, the Hamiltonian of a system that is otherwise quasi-Hermitian must cease to satisfy Eqs. (\ref{QH_def}) and (\ref{Dyson_def}) at an EP. That is to say, at an EP, quasi-Hermiticity and the associated unitary evolution therefore break down. Within the closed-system quasi-Hermitian interpretation adopted here, this is a quantum catastrophe \cite{Znojil_2012,Znojil_2020} and the probabilistic interpretation is lost \cite{Ghosh_2026}. There may nevertheless exist in the parameter space ``unitary corridors'' with quasi-Hermiticity that smoothly terminate at the EP  \cite{Znojil_2019,Znojil_2021,Bagchi_2026}. At such EPs, the spectral transition as the system passes from the quasi-Hermitian regime into a regime beyond the EP where no positive $\Theta$ satisfying Eq. (\ref{QH_def}) can be identified, has been interpreted in the literature as a phase transition \cite{Guo_2009,Bender_2013,Sergeev_2023}. 

\vspace{2mm}

On the other hand, fermionic chains are known to provide a setting to explore the manifestations of non-Hermiticity in many-body systems \cite{Li_2018,Hatano_1996,Hatano_1997,Prosen_2010,Ghatak_2018,Lee_2020,Mu_2020,Yang_2020,Gal_2023,Shi_2023,Kitahama_2024,Ma_2024a,Ardonne_2025,Zhang_2026,Jafari_2026,Pal_2026}. In particular, imbalanced pairing interactions that generalize those in the Kitaev chain \cite{Kitaev_2001} are known to produce intriguing features \cite{Li_2018,Ghatak_2018,Yang_2020,Shi_2023,Ma_2024a,Ardonne_2025,Zhang_2026,Jafari_2026}. Concerning EPs, the simplest two- and three-site cases display, respectively, EP2 and EP3 singularities \cite{Sinha_2024,Bagchi_2026}, motivating an explicit $N$-site realization for arbitrary $N$. At the general algebraic level, the fermionic lifting of a single-particle Jordan block to the many-body Fock space was analyzed in quadratic Liouville theory by Prosen \cite{Prosen_2010} and its Jordan decomposition was subsequently proven rigorously by Kitahama \textit{et al.} \cite{Kitahama_2024}. Related enhancement of the EP order for Kronecker-sum Hamiltonians was investigated in \cite{Wiersig_2025}. 

\vspace{2mm}

In this work we analyze an exactly solvable imbalanced $N$-site Kitaev chain, where a staggered particle-hole transformation reduces the problem to exterior-power representations of a tridiagonal one-defect matrix. We show that the one-defect sector achieves an EP with a size-$N$ Jordan block whose many-body lift realizes the Jordan amplification exposed in \cite{Prosen_2010,Kitahama_2024}, whereby the $m$-defect sector achieves the Jordan order $L_m=m(N-m)+1$, maximizing to $L_{\max}=\lfloor N^2/4\rfloor+1$. At this point all the $2^N$ many-body eigenvalues collapse to zero. We show that the EP arises as the terminal point of a quasi-Hermitian unitary corridor, which we explicitly realize using a positive metric \cite{Znojil_2019,Znojil_2021,Bagchi_2026}. In a distinct ``alternating-chain'' specialization, we show that the non-Hermitian pairing mechanism produces exact opposite-parity ground-state crossings within a quasi-Hermitian regime, and that their thermodynamic-limit accumulation generates a spectral Lifshitz transition \cite{MedinaCuy_2024,Lifshitz_1960} that is separated from the EP boundary. 

\vspace{2mm}

Our analysis therefore clearly separates the two kinds of spectral transitions, namely, the EP transition at which the Hamiltonian's diagonalizability is lost \cite{Guo_2009,Bender_2013} and the ground-state-crossing transition \cite{Sinha_2024} which we show exhibits signatures of quantum-critical behavior in the thermodynamic limit \cite{MedinaCuy_2024,Lifshitz_1960}. Thus our model of the exactly solvable imbalanced $N$-site Kitaev chain, which presents a precise many-body generalization of the lower-dimensional cases studied in \cite{Bagchi_2026,Sinha_2024}, provides a tractable model for understanding non-Hermitian criticality. 

\vspace{2mm}

The paper is organized as follows. In Sec. \ref{model_sec}, we introduce the model of the imbalanced $N$-site Kitaev chain and set up the basic notation. In Sec. \ref{invariant_jordanN_sec}, we expose an invariant-sector decomposition that yields, for $N\geq3$, two complementary sectors controlled by a one-defect tridiagonal matrix that develops size-$N$ Jordan blocks at eigenvalue coalescence. We then construct an explicit quasi-Hermitian corridor for this matrix that terminates at the EP. Sec. \ref{manybody_EP_order_sec} then lifts the one-body Jordan block to the full fermionic Fock space, revealing a many-body enhancement of the maximal Jordan order beginning at $N\geq4$. In Sec. \ref{parity_GS_sec}, we analyze the crossing between ground states of opposite fermion parity and show that its thermodynamic-limit accumulation produces Lifshitz-type criticality. Finally, we conclude with a discussion of the main results in Sec. \ref{disc_sec}. Some technical details are relegated to the Appendices \ref{appA}-\ref{appE}. 

\section{Theoretical model}\label{model_sec}
\begin{figure}[t]
\centering
\begin{tikzpicture}[x=1cm,y=1cm,font=\small]

  \node[anchor=west,font=\bfseries\small] at (-0.15,3.55)
    {(a) Imbalanced-pairing chain};

  \foreach \xpos/\siteindex in {0/1,1.55/2,3.10/3,4.65/4} {
    \node[qhsmallsite] (c\siteindex) at (\xpos,2.15)
      {$c_{\siteindex}$};
    \node[font=\scriptsize,below=1mm of c\siteindex]
      {$\varepsilon_{\siteindex}$};
  }

  \node at (5.45,2.15) {$\cdots$};

  \node[qhsmallsite] (cN) at (6.35,2.15) {$c_N$};
  \node[font=\scriptsize,below=1mm of cN]
    {$\varepsilon_N$};

  \draw[line width=0.85pt,draw=qhblue]
    (c1) to[bend left=18] (c2);
  \draw[line width=0.65pt,draw=qhgray,dashed]
    (c1) to[bend right=18] (c2);

  \foreach \siteL/\siteR in {c2/c3,c3/c4} {
    \draw[line width=0.85pt,draw=qhblue]
      (\siteL) to[bend left=16] (\siteR);
    \draw[line width=0.65pt,draw=qhgray,dashed]
      (\siteL) to[bend right=16] (\siteR);
  }

  \draw[line width=0.85pt,draw=qhblue]
    (c4) to[bend left=16] (5.15,2.15);
  \draw[line width=0.65pt,draw=qhgray,dashed]
    (c4) to[bend right=16] (5.15,2.15);

  \draw[line width=0.85pt,draw=qhblue]
    (5.75,2.15) to[bend left=16] (cN);
  \draw[line width=0.65pt,draw=qhgray,dashed]
    (5.75,2.15) to[bend right=16] (cN);

  \draw[line width=0.85pt,draw=qhblue]
    (0.15,3.02)--(0.51,3.02);
  \node[anchor=west,font=\scriptsize] at (0.58,3.02)
    {$\Delta_{j,+}c_j^\dagger c_{j+1}^\dagger$};

  \draw[line width=0.65pt,draw=qhgray,dashed]
    (3.42,3.02)--(3.78,3.02);
  \node[anchor=west,font=\scriptsize] at (3.85,3.02)
    {$\Delta_{j,-}c_{j+1}c_j$};

  \draw[qhmap]
    (3.15,1.25) -- (3.15,0.45)
    node[midway,right=2mm,align=left,font=\scriptsize]
    {$d_{2p-1}=c_{2p-1},\quad \quad d_{2p}=c_{2p}^{\dagger}$};

  \node[anchor=west,font=\bfseries\small] at (-0.15,0.25)
    {(b) Number-conserving defect chain};

  \foreach \xpos/\siteindex in {0/1,1.55/2,3.10/3,4.65/4} {
    \node[qhsmallsite] (d\siteindex) at (\xpos,-0.60)
      {$d_{\siteindex}$};
    \node[font=\scriptsize,below=1mm of d\siteindex]
      {$a_{\siteindex}$};
  }

  \node at (5.45,-0.60) {$\cdots$};

  \node[qhsmallsite] (dN) at (6.35,-0.60) {$d_N$};
  \node[font=\scriptsize,below=1mm of dN]
    {$a_N$};

  \foreach \siteL/\siteR/\bondindex
    in {d1/d2/1,d2/d3/2,d3/d4/3} {

    \draw[qhforward]
      ($(\siteL.east)+(0,0.09)$) --
      node[above=0.2mm,font=\scriptsize]
        {$v_{\bondindex}$}
      ($(\siteR.west)+(0,0.09)$);

    \draw[qhreverse]
      ($(\siteR.west)+(0,-0.09)$) --
      node[below=0.2mm,font=\scriptsize]
        {$u_{\bondindex}$}
      ($(\siteL.east)+(0,-0.09)$);
  }

  \draw[qhforward]
    ($(d4.east)+(0,0.09)$) -- (5.15,-0.51);
  \draw[qhreverse]
    (5.15,-0.69) -- ($(d4.east)+(0,-0.09)$);

  \draw[qhforward]
    (5.75,-0.51) -- ($(dN.west)+(0,0.09)$);
  \draw[qhreverse]
    ($(dN.west)+(0,-0.09)$) -- (5.75,-0.69);

\end{tikzpicture}

\caption{\justifying{The staggered particle-hole transformation on the even sublattice converts the two unequal pairing processes (solid blue and dashed gray) into number-conserving directed defect hopping. The onsite energies become $a_j=(-1)^{j+1}\varepsilon_j$. The right- and
left-directed hopping amplitudes are $K_{j+1,j}=v_j$ and
$K_{j,j+1}=u_j$, respectively.}}
\label{fig:model_defect_mapping}
\end{figure}
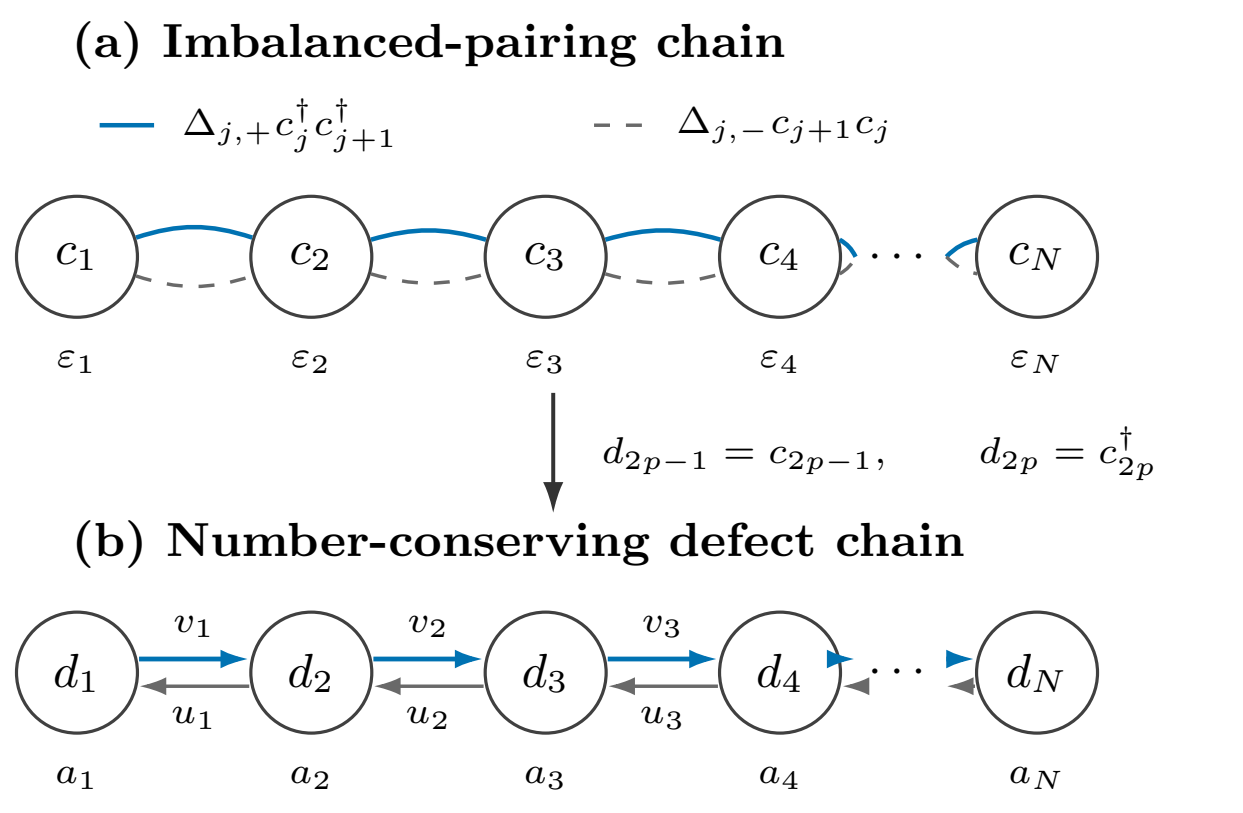

Our theoretical model is a one-dimensional chain of spinless fermions with $N\geq2$ sites [Fig. \ref{fig:model_defect_mapping}(a)] and the Hamiltonian
\begin{equation}
H = \sum_{j=1}^{N} \varepsilon_j
\left(n_j-\frac{1}{2}\right)
- \sum_{j=1}^{N-1} \left(\Delta_{j,+}
c_j^\dagger c_{j+1}^\dagger + \Delta_{j,-}
c_{j+1}c_j\right),
\label{imbalanced_kitaev}
\end{equation}
where $n_j=c_j^\dagger c_j$ is the number operator and $\varepsilon_j,\Delta_{j,\pm}\in\mathbb{R}$. Unless stated otherwise, we focus on nonzero, genuinely-imbalanced bonds $\Delta_{j,\pm}\neq0$ and $\Delta_{j,+}\neq\Delta_{j,-}$. The fermionic operators satisfy the canonical anticommutation relations
\begin{equation}
\{c_j,c_k^\dagger\}=\delta_{jk},
\quad \quad
\{c_j,c_k\}=\{c_j^\dagger,c_k^\dagger\}=0.
\label{canonical_anticommutation}
\end{equation}
The Hamiltonian features imbalanced $p$-wave pair-creation and pair-annihilation interactions between nearest neighbors, and so can be interpreted as a non-Hermitian Kitaev chain \cite{Kitaev_2001,Li_2018} in the absence of intersite hoppings. The common minus sign in Eq. (\ref{imbalanced_kitaev}) has no independent physical significance because the directed pairing amplitudes $\Delta_{j,+}$ and $\Delta_{j,-}$ are allowed to have either sign. For $N=2$ this corresponds to the fermionic extension of the Swanson oscillator \cite{Swanson_2004}, whose intriguing properties were studied in \cite{Sinha_2024}. The $N=3$ case was subsequently investigated in \cite{Bagchi_2026}, revealing EPs of the third order. While the quadratic nature of the model ensures its solvability, as we show, the model in Eq. (\ref{imbalanced_kitaev}) possesses several intriguing properties stemming from its non-Hermiticity.

\vspace{2mm}

We now introduce the number-generated similarity transformation
\begin{equation}
\Omega_>
=
\exp\left[
\sum_{j=1}^{N}
\theta_j
\left(
n_j-\frac{1}{2}
\right)
\right],
\quad \quad
\theta_j\in{\mathbb R},
\label{pairing_dyson_map}
\end{equation}
and for finite real $\theta_j$, the operator $\Omega_>$ is Hermitian, strictly positive, and invertible, and so is the underlying metric $\Theta_> = \Omega_>^2$. With this the Hermitian representative $h=\Omega_> H\Omega_>^{-1}$ becomes
\begin{equation}
h=
\sum_{j=1}^{N}
\varepsilon_j
\left(
n_j-\frac{1}{2}
\right)
-
\sum_{j=1}^{N-1}
\overline{\Delta}_j
\left(
c_j^\dagger c_{j+1}^\dagger
+
c_{j+1}c_j
\right),
\label{hermitian_pairing_chain}
\end{equation}
where the reciprocal pairing amplitudes $\overline{\Delta}_j$ are real and nonzero, and with $g_j=\theta_j+\theta_{j+1}$, one has $\Delta_{j,+} = \overline{\Delta}_j e^{-g_j}$ and $\Delta_{j,-} = \overline{\Delta}_j e^{g_j}$. Such a Dyson map works on every bond precisely when
\begin{equation}
\Delta_{j,+}\Delta_{j,-}>0,
\quad \quad
j=1,\cdots,N-1.
\label{positive_bond_product}
\end{equation}
Indeed, the inverse parameter map is
\begin{eqnarray}
\overline{\Delta}_j
&=&
{\rm sgn}(\Delta_{j,+})
\sqrt{\Delta_{j,+}\Delta_{j,-}},
\nonumber\\
g_j
&=&
\frac{1}{2}
\ln\left(
\frac{\Delta_{j,-}}{\Delta_{j,+}}
\right).
\label{inverse_parameter_map}
\end{eqnarray}
Because the chain is open, the $N-1$ conditions
$\theta_j+\theta_{j+1}=g_j$ always admit a real solution, given by
\begin{equation}
\theta_j
=
(-1)^{j-1}\theta_1
+
\sum_{k=1}^{j-1}
(-1)^{j-1-k}g_k.
\label{theta_general_solution}
\end{equation}
This construction of our model in Eq. (\ref{imbalanced_kitaev}) exposes the fact that the model, while non-Hermitian for $\Delta_{j,+} \neq \Delta_{j,-}$, is quasi-Hermitian with the Dyson map implemented by $\Omega_>$ when Eq. (\ref{positive_bond_product}) holds. This therefore excludes the possibility of EPs for positive bond products in Eq. (\ref{positive_bond_product}). Our original model in Eq. (\ref{imbalanced_kitaev}), however, allows also for 
\begin{equation}
    \Delta_{j,+}\Delta_{j,-}<0,
\label{nontrivial_condition}
\end{equation} 
on one or more bonds. In that case no real choice of the density-generated parameters $\theta_j$ can work because $\overline{\Delta}_j^{2}$ would be negative. Nevertheless, the failure of this local construction does not automatically exclude quasi-Hermiticity implemented by a similarity transformation that differs from Eq. (\ref{pairing_dyson_map}), though it opens up the question regarding the existence of EPs in the regime of Eq. (\ref{nontrivial_condition}). As we show in subsequent sections, the regime of Eq. (\ref{nontrivial_condition}) can indeed admit EPs and in particular, we will construct an explicit negative-product quasi-Hermitian subfamily and its EP endpoint in Sec. \ref{sec:negative_product_Dyson_map} for every $N\geq2$. Exactly at its terminal point, diagonalizability is lost and no positive-definite invertible metric can survive.

\section{Eigenvalue collapse and one-defect Jordan block}\label{invariant_jordanN_sec}

\subsection{Invariant-sector decomposition}
\label{subsec:N_site_decomposition}

The finite-dimensional reduction used in the two- and three-site fermionic Swanson systems \cite{Bagchi_2026,Sinha_2024} extends to arbitrary finite $N$. In fact, the model in Eq. (\ref{imbalanced_kitaev}) possesses a conserved charge, defined by the staggered number operator 
\begin{equation}
\mathcal{Q} =
\sum_{j=1}^{N}
(-1)^{j+1}
\left(
n_j-\frac{1}{2}
\right).
\label{staggered_charge}
\end{equation}
The origin of this conservation law is particularly transparent: as pair creation or annihilation changes the occupations of two neighboring sites simultaneously, the two sites enter Eq. (\ref{staggered_charge}) with opposite signs. More explicitly, one has
\begin{equation}
[\mathcal{Q},c_j^\dagger c_{j+1}^\dagger] = [\mathcal{Q},
c_{j+1}c_j] = 0,
\label{staggered_charge_commutator}
\end{equation}
giving therefore $[\mathcal{Q},H]=0$. It is useful to introduce the binary defect occupations
\begin{equation}
\nu_j
=
\frac{1}{2}
+
(-1)^{j+1}
\left(
n_j-\frac{1}{2}
\right)
=
\left\{
\begin{array}{ll}
n_j, & j\ {\rm odd},\\[1mm]
1-n_j, & j\ {\rm even}.
\end{array}
\right.
\label{defect_occupations}
\end{equation}
In terms of these variables, one has
\begin{equation}
{\cal Q}
=
\sum_{j=1}^{N}\nu_j-\frac{N}{2},
\label{staggered_charge_defects}
\end{equation}
meaning that if $m=\sum_j\nu_j$, the eigenvalue of $\mathcal{Q}$ is simply $m-N/2$. Now there are $\binom{N}{m}$ independent defect configurations with fixed $m$, and so the complete Fock-space Hamiltonian decomposes as
\begin{equation}
H \cong \bigoplus_{m=0}^{N}H_m, \quad \quad \dim H_m=\binom{N}{m}.
\label{N_site_direct_sum}
\end{equation}
Every such sector also has a definite conventional fermion parity since one can write
\begin{equation}
(-1)^{\sum_j n_j}
=
(-1)^{m+\lfloor N/2\rfloor},
\label{defect_sector_parity}
\end{equation}
where the symbol $\lfloor x \rfloor$ denotes the greatest integer less than or equal to $x$. The tantalizing role of fermion parity will be discussed explicitly in Sec. \ref{parity_GS_sec}. The staggered-charge decomposition is nevertheless finer than the ordinary even-odd parity decomposition and is the appropriate generalization of the finite-matrix partitioning employed in \cite{Sinha_2024,Bagchi_2026}. For $N=2$, Eq. (\ref{N_site_direct_sum}) gives sector dimensions $(1,2,1)$, with the two-dimensional sector containing the vacuum and the doubly occupied state \cite{Sinha_2024}. For $N=3$, it gives sector dimensions $(1,3,3,1)$, thereby reproducing the two isolated levels and the two nontrivial three-dimensional blocks \cite{Bagchi_2026}.

\subsection{Exceptional point of order $N$ in the $m=1, N-1$ sectors}
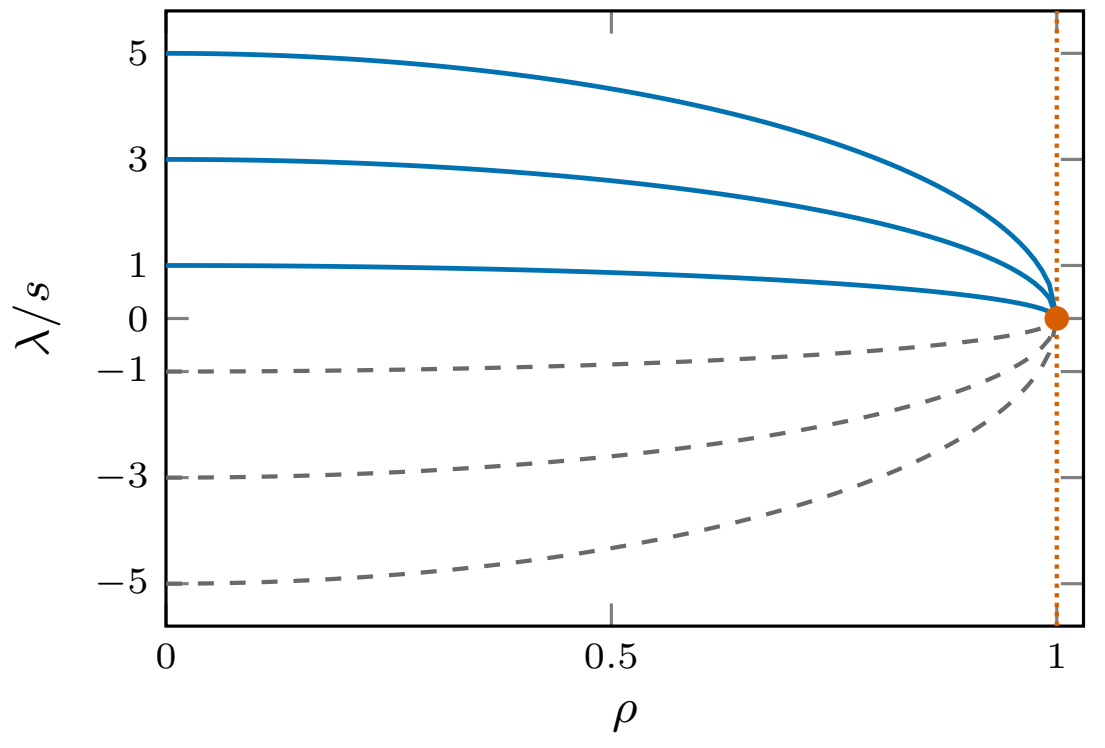
\begin{figure}[t]
\centering
\begin{tikzpicture}
  \begin{axis}[
    width=7.65cm,
    height=5.65cm,
    xmin=0,
    xmax=1.03,
    ymin=-5.8,
    ymax=5.8,
    xlabel={$\rho$},
    ylabel={$\lambda/s$},
    xtick={0,0.5,1},
    ytick={-5,-3,-1,0,1,3,5},
    tick label style={font=\scriptsize},
    label style={font=\small},
    axis line style={line width=0.6pt},
    tick style={line width=0.6pt},
    domain=0:1,
    samples=121,
    clip=false
  ]

    \addplot[qhblue,line width=0.9pt] {5*sqrt(max(0,1-x^2))};
    \addplot[qhblue,line width=0.9pt] {3*sqrt(max(0,1-x^2))};
    \addplot[qhblue,line width=0.9pt] {sqrt(max(0,1-x^2))};

    \addplot[qhgray,dashed,line width=0.8pt] {-sqrt(max(0,1-x^2))};
    \addplot[qhgray,dashed,line width=0.8pt] {-3*sqrt(max(0,1-x^2))};
    \addplot[qhgray,dashed,line width=0.8pt] {-5*sqrt(max(0,1-x^2))};

    \draw[qhred,densely dotted,line width=0.8pt]
      (axis cs:1,-5.8)--(axis cs:1,5.8);

    \addplot[qhred,only marks,mark=*,mark size=2.1pt]
      coordinates {(1,0)};

  \end{axis}
\end{tikzpicture}

\caption{\justifying{One-defect spectral collapse along the
quasi-Hermitian corridor for $N=6$. The levels in
Eq. (\ref{negative_product_path_spectrum}) are real and distinct
for $0\leq\rho<1$. At $\rho=1$, all six levels coalesce at zero
and $K_{6,{\rm EP}}\sim J_6(0)$.}}
\label{fig:qh_endpoint_spectrum}
\end{figure}

The sectors $m=1$ and $m=N-1$ are of particular importance because both of them have dimension $N$ and hence may be able to support a Jordan block of size $N$. The sector decomposition becomes especially transparent after a staggered particle-hole transformation given by
\begin{equation}
d_j
=
\left\{
\begin{array}{ll}
c_j, & j\ {\rm odd},\\[1mm]
c_j^\dagger, & j\ {\rm even}.
\end{array}
\right.
\quad \quad
d_j^\dagger d_j=\nu_j,
\label{defect_fermions}
\end{equation}
where $d_j$ obey the canonical anticommutation relations. The operators $d_j$ implement a particle-hole transformation on the even sublattice, converting every pairing bond in Eq.
(\ref{imbalanced_kitaev}) into number-conserving defect hopping. We refer to the resulting particles and holes, measured relative to the alternating reference configuration, as defect fermions. The defect vacuum $|\Lambda\rangle$, defined by $d_j|\Lambda\rangle=0$, has the occupation pattern $(0,1,0,1,\cdots)$. Introducing
\begin{equation}
a_j=(-1)^{j+1}\varepsilon_j,
\quad \quad
E_\Lambda
=
-\frac{1}{2}\sum_{j=1}^{N}a_j,
\label{defect_onsite_energy}
\end{equation}
the Hamiltonian in Eq. (\ref{imbalanced_kitaev}) can be expressed exactly as
\begin{equation}
H
=
E_\Lambda
+
\sum_{j,k=1}^{N}
(K_N)_{jk}d_j^\dagger d_k,
\label{defect_number_conserving_form}
\end{equation}
where $K_N$ is of the following tridiagonal form: 
\begin{equation}
K_N=
\left(
\begin{array}{ccccc}
a_1 & u_1 & & & \\
v_1 & a_2 & u_2 & & \\
& v_2 & a_3 & \ddots & \\
& & \ddots & \ddots & u_{N-1}\\
& & & v_{N-1} & a_N
\end{array}
\right),
\label{KN_definition}
\end{equation}
with
\begin{eqnarray}
u_j
&=&
\left\{
\begin{array}{ll}
-\Delta_{j,+}, & j\ {\rm odd},\\
\phantom{-}\Delta_{j,-}, & j\ {\rm even},
\end{array}
\right.
\nonumber\\
v_j
&=&
\left\{
\begin{array}{ll}
-\Delta_{j,-}, & j\ {\rm odd},\\
\phantom{-}\Delta_{j,+}, & j\ {\rm even}.
\end{array}
\right.
\label{KN_offdiagonal}
\end{eqnarray}
Consequently, one has
\begin{equation}
u_jv_j = \Delta_{j,+}\Delta_{j,-},
\label{bond_product_invariant}
\end{equation}
independent of the basis-sign convention. The staggered particle-hole map and its conversion of imbalanced pairing into directed, number-conserving defect hopping are illustrated in Fig. \ref{fig:model_defect_mapping}. The restriction to the one-defect sector is therefore
\begin{equation}
H_1=E_\Lambda I_N+K_N.
\label{one_defect_block}
\end{equation}
Letting $\mathcal{K}_k$ denote the leading $k\times k$ principal block of $K_N$ and defining $P_k(E)=\det(EI_k-\mathcal{K}_k)$, one has
\begin{eqnarray}
P_0(E)&=&1, \label{general_determinant_recursion}\\
P_1(E)&=&E-a_1,
\nonumber\\
P_k(E)
&=&
(E-a_k)P_{k-1}(E)
-
u_{k-1}v_{k-1}P_{k-2}(E). \nonumber
\end{eqnarray}
Using Eqs. (\ref{defect_onsite_energy}) and
(\ref{bond_product_invariant}), this becomes
\begin{eqnarray}
P_k(E)
&=&
\left[
E-(-1)^{k+1}\varepsilon_k
\right]
P_{k-1}(E)
\nonumber\\
&&
-
\Delta_{k-1,+}\Delta_{k-1,-}
P_{k-2}(E).
\label{physical_determinant_recursion}
\end{eqnarray}
In particular, $P_N(E)$ is the characteristic polynomial of $K_N$. Eq. (\ref{physical_determinant_recursion}) shows that the two directed amplitudes on each bond enter the spectrum through their product $u_jv_j=\Delta_{j,+}\Delta_{j,-}$. Now we take the parametrization
\begin{eqnarray}
a_j=s(N+1-2j), \quad \quad
u_jv_j=-s^2\rho^2j(N-j),
\label{negative_product_path}
\end{eqnarray}
where $s\in\mathbb{R}\setminus\{0\}$ and $0\leq\rho\leq1$. A realization with uniformly bounded local parameters is obtained by taking $s=s_0/N$ with fixed $s_0\ne0$ and choosing $\Delta_{j,+}=-\Delta_{j,-}=-s\rho \sqrt{j(N-j)}$. For $0\leq\rho<1$, the eigenvalues of $K_N$ are (see proof in Sec. \ref{sec:negative_product_Dyson_map})
\begin{equation}
\lambda_j(\rho)=s(N+1-2j)\sqrt{1-\rho^2},
\quad \quad j=1,\cdots,N.
\label{negative_product_path_spectrum}
\end{equation}
The level spacing is $2|s|\sqrt{1-\rho^2}$, so all the $N$ levels reach zero at $\rho=1$, as shown in Fig. \ref{fig:qh_endpoint_spectrum}. 
In terms of the physical parameters appearing in the Hamiltonian in Eq. (\ref{imbalanced_kitaev}), the coalescence condition follows from Eq. (\ref{negative_product_path}) to be
\begin{eqnarray}
(-1)^{j+1}\varepsilon_j
&=&s(N+1-2j),\quad \quad j=1,\cdots,N,
\label{EPN_parameter_family}\\
\Delta_{j,+}\Delta_{j,-}
&=&-s^2j(N-j),\quad \quad j=1,\cdots,N-1.
\nonumber
\end{eqnarray}
In this case the characteristic polynomial reads
\begin{equation}
P_N(E)=E^N.
\label{N_fold_degeneracy}
\end{equation}
All $u_j$ are nonzero, so the first $N-1$ rows of
$K_N\psi=0$ determine $\psi_2,\cdots,\psi_N$ from $\psi_1$. The zero eigenspace is therefore one-dimensional and we denote this defective matrix as $K_{N,{\rm EP}}\equiv K_N(1)$, so that $K_{N,{\rm EP}}\sim J_N(0)$ (see Appendix \ref{appA} for rigorous proof). Since $E_\Lambda=0$, this is also the Jordan form of $H_1$. One can similarly show that the $m=N-1$ sector admits a size-$N$ Jordan block at the EP. The sectors $m=1$ and $m=N-1$ coincide for $N=2$ and are distinct for $N\geq3$.

\subsection{Quasi-Hermitian corridor and its endpoint}
\label{sec:negative_product_Dyson_map}

We now show explicitly that the EP can be approached continuously from a quasi-Hermitian regime. For this purpose it is sufficient to consider one simple
realization of the parameter path in Eq. (\ref{negative_product_path}) with $\Delta_{j,+}=-\Delta_{j,-}=-s\rho \sqrt{j(N-j)}$. This immediately gives $\Delta_{j,+}\Delta_{j,-}=-s^2\rho^2j(N-j)$, which is strictly negative for $0<\rho\leq1$ and hence satisfies Eq. (\ref{nontrivial_condition}), with the Hamiltonian becoming Hermitian at $\rho=0$. Using Eq. (\ref{KN_offdiagonal}), one finds, independently of whether $j$ is even or odd, that
\begin{equation}
u_j=s\rho\sqrt{j(N-j)},\qquad
v_j=-s\rho\sqrt{j(N-j)}.
\label{negative_product_simple_hopping}
\end{equation}
Letting $J=(N-1)/2$, we introduce the spin-$J$ matrices through
\begin{equation}
(J_z)_{jj}
=
\frac{N+1-2j}{2},
\qquad
(J_+)_{j,j+1}
=
\sqrt{j(N-j)},
\label{spin_matrices_definition}
\end{equation}
with $J_-=J_+^\dagger$, $2J_x=J_++J_-$, and
$2iJ_y=J_+-J_-$. Comparing Eqs. (\ref{KN_definition}),
(\ref{negative_product_path}), and
(\ref{negative_product_simple_hopping}), the one-defect matrix immediately becomes
\begin{equation}
K_N(\rho)
=
2s\left(J_z+i\rho J_y\right).
\label{negative_product_spin_form}
\end{equation}
In order to expose quasi-Hermiticity as well as obtain the spectrum stated in Eq. (\ref{negative_product_path_spectrum}), for $0\leq\rho<1$ we introduce
\begin{equation}
\eta=\operatorname{arctanh}\rho,
\qquad
\Omega_\rho^{(1)}=e^{\eta J_x}.
\label{one_defect_spin_dyson}
\end{equation}
Now the angular-momentum commutators
$[J_x,J_z]=-iJ_y$ and $[J_x,iJ_y]=-J_z$ give
\begin{eqnarray}
e^{\eta J_x}J_ze^{-\eta J_x}
&=&
J_z\cosh\eta-iJ_y\sinh\eta,
\nonumber\\
e^{\eta J_x}(iJ_y)e^{-\eta J_x}
&=&
iJ_y\cosh\eta-J_z\sinh\eta.
\label{spin_hyperbolic_rotation}
\end{eqnarray}
Applying Eq. (\ref{spin_hyperbolic_rotation}) to
Eq. (\ref{negative_product_spin_form}) yields
\begin{eqnarray}
&&e^{\eta J_x}
\left(J_z+i\rho J_y\right)
e^{-\eta J_x}
\nonumber\\
&&\quad =
J_z\left(\cosh\eta-\rho\sinh\eta\right) \nonumber\\
&&\quad +
iJ_y\left(\rho\cosh\eta-\sinh\eta\right).
\label{spin_rotation_intermediate}
\end{eqnarray}
Since $\tanh\eta=\rho$, one has
$\sinh\eta=\rho\cosh\eta$ and therefore the coefficient of $J_y$ in Eq. (\ref{spin_rotation_intermediate}) vanishes. Moreover, $\cosh\eta=(1-\rho^2)^{-1/2}$, so we have $\cosh\eta-\rho\sinh\eta=\sqrt{1-\rho^2}$ and therefore one gets the Dyson map
\begin{equation}
\Omega_\rho^{(1)}
K_N(\rho)
\left(\Omega_\rho^{(1)}\right)^{-1} =
2s\sqrt{1-\rho^2} J_z.
\label{negative_product_Hermitian_similarity}
\end{equation}
A closely related complexified spin transformation was employed in a noninteracting non-Hermitian Bose-Hubbard model \cite{Graefe_2008}. Eq. (\ref{negative_product_Hermitian_similarity}) shows that $K_N(\rho)$ is similar to the Hermitian matrix $2s\sqrt{1-\rho^2} J_z$ and so must have real eigenvalues. In order to connect this explicitly with the quasi-Hermiticity condition in Eq. (\ref{QH_def}), we let
\begin{equation}
h_\rho^{(1)}
=
\Omega_\rho^{(1)}
K_N(\rho)
\left(\Omega_\rho^{(1)}\right)^{-1}
=
2s\sqrt{1-\rho^2} J_z.
\label{one_defect_Hermitian_representative}
\end{equation}
Since $h_\rho^{(1)}=(h_\rho^{(1)})^\dagger$, taking the Hermitian
conjugate of Eq. (\ref{one_defect_Hermitian_representative}) and
comparing the two expressions gives
\begin{equation}
K_N^\dagger(\rho) 
\Theta_\rho^{(1)}
=
\Theta_\rho^{(1)}K_N(\rho),
\qquad
\Theta_\rho^{(1)}
=
\left(\Omega_\rho^{(1)}\right)^\dagger
\Omega_\rho^{(1)}.
\label{negative_product_one_defect_qh}
\end{equation}
For $0\leq\rho<1$, $\eta$ is finite and since
$J_x=J_x^\dagger$, the operator $\Omega_\rho^{(1)}=e^{\eta J_x}$ is Hermitian, strictly positive, and invertible. Therefore we conclude that 
\begin{equation}
\Theta_\rho^{(1)}
=
e^{2\eta J_x}>0,
\label{negative_product_one_defect_metric}
\end{equation}
and Eq. (\ref{negative_product_one_defect_qh}) is equivalently $K_N^\dagger=\Theta_\rho^{(1)}K_N(\Theta_\rho^{(1)})^{-1}$. Thus $K_N(\rho)$ is quasi-Hermitian throughout $0\leq\rho<1$. The extension of this construction to the complete many-body Hamiltonian is given in Appendix \ref{appB}.

\vspace{2mm}

Finally, we expose the spectrum quoted in Eq. (\ref{negative_product_path_spectrum}). Referring to Eq. (\ref{negative_product_Hermitian_similarity}), since similarity transformations preserve eigenvalues while $J_z$ is diagonal with eigenvalues $(N+1-2j)/2$, the spectrum in Eq. (\ref{negative_product_path_spectrum}) directly follows, with eigenvalues being  real and distinct for every $0\leq\rho<1$. As
$\rho\rightarrow1^{-}$, all these eigenvalues collapse to zero while $\eta=\operatorname{arctanh}\rho\rightarrow\infty$, so the similarity transformation becomes singular, as expected at the EP. The interval $0\leq\rho<1$ thus provides a quasi-Hermitian corridor whose terminal point is the EP at $\rho=1$.

\section{Many-body enhancement of the maximal Jordan order}
\label{manybody_EP_order_sec}

So far we have focused on the $m=1$ and $m=N-1$ sectors, which admit size-$N$ Jordan blocks at the EP. At the general algebraic level, the fermionic lift of a principal one-defect block is known to have largest Jordan-block size $m(N-m)+1$ in $\bigwedge^m{\mathbb C}^N$ \cite{Prosen_2010,Kitahama_2024}. We now show that our model explicitly realizes this, using the induced Lie-algebra action $d\Gamma_m[J_N(0)]$, which should be distinguished from the multiplicative exterior power $\bigwedge^m J_N(0)$.

\subsection{Jordan blocks of size larger than $N$}

Because Eq. (\ref{defect_number_conserving_form}) is exactly number conserving in the defect fermions, its restriction to the $m$-defect sector is the standard induced one-defect representation on $\bigwedge^m{\mathbb C}^N$. In particular, restricting Eq. (\ref{defect_number_conserving_form}) to the $m$-defect sector gives
\begin{equation}
H_m
=
E_\Lambda I_{\binom{N}{m}}
+
d\Gamma_m(K_N),
\label{many_body_second_quantization}
\end{equation}
where $d\Gamma_m(K_N)$ is the generator induced on
$\bigwedge^m{\mathbb C}^N$, explicitly given by (see Appendix \ref{appC})
\begin{eqnarray}
d\Gamma_m(K_N)
\left(
w_1\wedge\cdots\wedge w_m
\right)
&=&
\sum_{a=1}^{m}
w_1\wedge\cdots
\label{induced_many_body_generator} \\
&&
\wedge K_Nw_a
\wedge\cdots\wedge w_m.
\nonumber
\end{eqnarray} 
At the principal EP tuning, one may write
$K_{N,{\rm EP}}=S_NJ_N(0)S_N^{-1}$, and the functoriality of the exterior representation then gives
\begin{equation}
d\Gamma_m(K_{N,{\rm EP}})
=
\left(\bigwedge\nolimits^m S_N\right)
d\Gamma_m[J_N(0)]
\left(\bigwedge\nolimits^m S_N\right)^{-1}.
\label{many_body_similarity}
\end{equation}
At this tuning, we have $E_\Lambda=0$ and we can therefore denote the restriction to the $m$-defect sector and the complete Fock-space Hamiltonian by
\begin{eqnarray}
H_{m,{\rm EP}}
&\equiv&
d\Gamma_m(K_{N,{\rm EP}}), \nonumber\\
H_{\rm EP}
&\equiv&
d\Gamma(K_{N,{\rm EP}})
\cong
\bigoplus_{m=0}^{N}H_{m,{\rm EP}}.
\label{EP_sector_decomposition}
\end{eqnarray}
The Jordan structure of every sector is consequently determined completely by
$d\Gamma_m[J_N(0)]$, without involving any additional parameter tuning.

\vspace{2mm}

To determine the Jordan-chain lengths, let us choose a Jordan basis $\{e_0,e_1,\cdots,e_{N-1}\}$ such that $J_N(0)e_i=e_{i-1}$, with $e_{-1}=0$, and define
\begin{equation}
|i_1,\cdots,i_m\rangle
=
e_{i_1}\wedge\cdots\wedge e_{i_m},
\quad \quad
0\leq i_1<\cdots<i_m\leq N-1.
\label{exterior_basis}
\end{equation}
Writing $T_m=d\Gamma_m[J_N(0)]$, its action is
\begin{eqnarray}
T_m|i_1,\cdots,i_m\rangle
&=&
\sum_{a=1}^{m}
e_{i_1}\wedge\cdots
\nonumber\\
&&
\wedge e_{i_a-1}
\wedge\cdots\wedge e_{i_m}.
\label{Tm_lowering_action}
\end{eqnarray}
Terms with $i_a=0$ vanish, as do terms for which
$i_a-1=i_{a-1}$, because the corresponding wedge product contains two identical vectors. Now let us define the nonnegative grade
\begin{equation}
G(i_1,\cdots,i_m)
=
\sum_{a=1}^{m}
\left[
i_a-(a-1)
\right],
\label{wedge_grade}
\end{equation}
so that every nonvanishing action of $T_m$ lowers this grade by exactly one. The minimum is attained at $(0,1,\cdots,m-1)$, whereas the maximum is attained at $(N-m,\cdots,N-1)$, i.e., one can write
\begin{equation}
G_{\rm min}=0, \quad \quad G_{\rm max}=m(N-m),
\label{wedge_grade_difference}
\end{equation}
and therefore it follows that $T_m^{m(N-m)+1}=0$. This upper bound is saturated. Starting from
$|N-m,\cdots,N-1\rangle$, one may first lower its smallest occupied index to $0$, then the next occupied index to $1$, and continue until the final configuration $|0,\cdots,m-1\rangle$ is reached. This produces an allowed path containing exactly $m(N-m)$ lowering steps. Moreover, in the ordered exterior basis, every allowed elementary lowering has a positive coefficient, i.e., lowering an index by one
preserves its position in the ordered wedge unless a collision occurs, in which case the term vanishes. Hence different allowed paths contribute with the same sign and cannot cancel. As a consequence, one has
\begin{equation}
\left\langle
0,\cdots,m-1
\right|
T_m^{m(N-m)}
\left|
N-m,\cdots,N-1
\right\rangle
>0,
\label{Tm_nonzero_matrix_element}
\end{equation}
and therefore $T_m^{m(N-m)}|N-m,\cdots,N-1\rangle\neq0$. The nilpotency index, and hence the size of the largest Jordan block
in the $m$-defect sector, is therefore
\begin{equation}
L_m=m(N-m)+1.
\label{many_body_sector_EP_order}
\end{equation}
In other words, $H_{m,{\rm EP}}$ contains a Jordan block of dimension $L_m$ and no larger Jordan block can occur in that sector. Since every $H_{m,{\rm EP}}$ is nilpotent, all $2^N$ many-body eigenvalues collapse to zero at the EP. For $m=1$, Eq. (\ref{many_body_sector_EP_order}) gives $L_1=N$ while the particle-hole complementarity gives $L_{N-1}=N$.

\subsection{Jordan block of maximal size}
The enhancement is the largest near $m=N/2$ in the defect
representation, where the number of allowed fermionic lowering
operations is maximal. Maximizing Eq. (\ref{many_body_sector_EP_order}) over the integer
$m$ gives
\begin{equation}
L_{\rm max} = \max_{0\leq m\leq N}m(N-m) + 1
= \left\lfloor
\frac{N^2}{4}
\right\rfloor+1.
\label{maximal_many_body_EP_order}
\end{equation}
When several Jordan blocks occur at the same coalesced eigenvalue, we call the size of the largest block the \emph{maximal Jordan order}. The numbers $L_m$ and $L_{\rm max}$ are maximal Jordan orders and the complete Hamiltonian is generally a direct sum.

\vspace{2mm}

\begin{figure*}[t]
\centering
\begin{tikzpicture}

  \begin{axis}[
    name=sectororder,
    at={(0cm,0cm)},
    anchor=south west,
    width=7.70cm,
    height=5.55cm,
    xmin=-0.35,
    xmax=8.35,
    ymin=0,
    ymax=19,
    xlabel={$m$},
    ylabel={$L_m$},
    xtick={0,1,...,8},
    ytick={0,4,8,12,16},
    tick label style={font=\scriptsize},
    label style={font=\small},
    axis line style={line width=0.6pt},
    tick style={line width=0.6pt},
    clip=false
  ]

    \addplot[
      qhblue,
      ycomb,
      line width=0.85pt,
      mark=*,
      mark size=2.0pt,
      mark options={solid,fill=white}
    ]
      coordinates {
        (0,1)
        (1,8)
        (2,13)
        (3,16)
        (4,17)
        (5,16)
        (6,13)
        (7,8)
        (8,1)
      };

    \addplot[
      qhorange,
      densely dashed,
      line width=0.85pt,
      domain=0:8,
      samples=81
    ]
      {x*(8-x)+1};

    \node[
      anchor=north west,
      font=\bfseries
    ]
      at (axis description cs:0.025,0.975)
      {(a)};

  \end{axis}

  \begin{axis}[
    name=maxorder,
    at={(8.45cm,0cm)},
    anchor=south west,
    width=7.70cm,
    height=5.55cm,
    xmin=1.6,
    xmax=12.4,
    ymin=0,
    ymax=40,
    xlabel={$N$},
    ylabel={$L_{\mathrm{max}}$},
    xtick={2,4,6,8,10,12},
    ytick={0,10,20,30,40},
    tick label style={font=\scriptsize},
    label style={font=\small},
    axis line style={line width=0.6pt},
    tick style={line width=0.6pt},
    clip=false
  ]

    \addplot[
      qhblue,
      line width=0.95pt,
      mark=*,
      mark size=2.0pt,
      mark options={solid,fill=white}
    ]
      coordinates {
        (2,2)
        (3,3)
        (4,5)
        (5,7)
        (6,10)
        (7,13)
        (8,17)
        (9,21)
        (10,26)
        (11,31)
        (12,37)
      };

    \addplot[
      qhgray,
      dashed,
      line width=0.85pt,
      domain=2:12
    ]
      {x};

    \node[
      anchor=north west,
      font=\bfseries
    ]
      at (axis description cs:0.025,0.975)
      {(b)};

  \end{axis}

\end{tikzpicture}

\caption{\justifying{Sector-resolved and maximal many-body Jordan orders. (a) For $N=8$, the blue markers and vertical stems show the largest Jordan block size $L_m=m(8-m)+1$ in each integer defect sector $m=0,\cdots,8$. The maximum is $L_4=17$ in the half-filled defect sector. (b) The blue markers show $L_{\mathrm{max}}=\lfloor N^2/4\rfloor+1$ as a function of chain length, whereas the dashed gray line shows the one-defect order $N$. The maximal many-body order first exceeds $N$ at $N=4$ and grows quadratically for large $N$.}}
\label{fig:jordan_hierarchy_scaling}
\end{figure*}
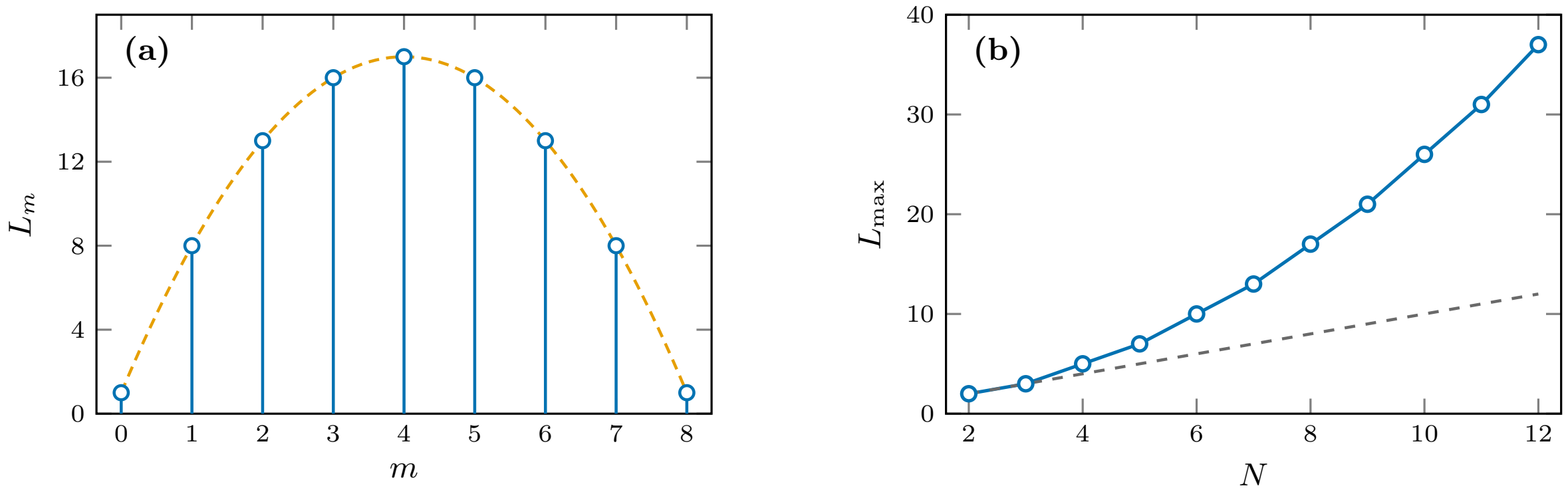

For even $N$, the maximum occurs uniquely at $m=N/2$, whereas for odd $N$ it occurs in the two complementary sectors $m=(N-1)/2$ and $m=(N+1)/2$. The maximal Jordan-chain length of the many-body Hamiltonian therefore scales quadratically with the number of
sites in the manner
\begin{equation}
L_{\rm max}
\sim
\frac{N^2}{4},
\quad \quad
N\gg1.
\label{quadratic_EP_scaling}
\end{equation}
This quadratic growth should be contrasted with the exponential growth, $2^N$, of the full Fock-space dimension. The sector-resolved Jordan hierarchy and the scaling of its maximal order are summarized in Fig. \ref{fig:jordan_hierarchy_scaling}. The enhancement has a simple fermionic interpretation, that is, Pauli exclusion permits each of the $m$ defects in the longest chain to undergo $N-m$ net Jordan-level lowerings, giving $m(N-m)$ steps and hence Eq. (\ref{many_body_sector_EP_order}).

\section{Parity-protected ground-state crossings and Lifshitz criticality}\label{parity_GS_sec}

Apart from the EP, we now show that an intriguing feature of our model is the presence of ground-state crossings at finite $N$. In the following, the phrase ``parity protected'' will be used to imply that fermion-parity conservation forbids any hybridization between the two opposite-parity levels when they meet. The simplest case for $N=2$ was observed in \cite{Sinha_2024} and we now generalize the result to every finite $N$. 

\subsection{Parity protection of the ground-state crossing}
\label{subsec:general_ground_state_crossing}

The staggered particle-hole transformation of Eq. (\ref{defect_fermions}) converts the conserved charge into the defect number $\widehat{M}=\sum_jd_j^\dagger d_j={\cal Q}+N/2$ and casts the Hamiltonian in the number-conserving form of Eq. (\ref{defect_number_conserving_form}). We focus on the regime where the one-defect matrix $K_N$ is quasi-Hermitian, i.e., it is diagonalizable and has real eigenvalues $\{\lambda_\mu\}_{\mu=1}^{N}$. Its diagonalization induces a biorthogonal canonical transformation of the defect fermions. Consequently, the many-body energies are obtained by occupying distinct one-defect levels and the ground-state energy is
\begin{equation}
E_{\rm gs}
=
E_\Lambda+
\sum_{\mu=1}^{N}\min(0,\lambda_\mu).
\label{general_ground_state_energy}
\end{equation}
While $\lambda_\mu$ are the one-defect eigenvalues, a many-body state with mode occupations $f_\mu\in\{0,1\}$ has the energy $E=E_\Lambda+\sum_\mu f_\mu\lambda_\mu$. Eq. (\ref{general_ground_state_energy}) minimizes over all occupation patterns $\{f_\mu\}$, hence over all defect numbers $m=\sum_\mu f_\mu$. Within a fixed physical-parity sector, the occupations must additionally satisfy the corresponding parity constraint. When we compare states from all defect sectors, $f_\mu=1$ for $\lambda_\mu<0$ and $f_\mu=0$ for $\lambda_\mu>0$, so away from zero modes the ground-state defect number is $M_{\rm gs}=\sum_\mu f_\mu=\#\{\mu:\lambda_\mu<0\}$. Let $\lambda_{\rm cr}$ denote a particular algebraically simple eigenvalue of $K_N$ that changes sign as a control parameter is varied while all other one-defect eigenvalues remain nonzero at its crossing. Its occupation $f_{\rm cr}$ then switches between $0$ and $1$, changing $M_{\rm gs}$ by one.

\vspace{2mm}

Let $x$ denote some controllable parameter so that the characteristic polynomial is expressible as $P_N(E;x)=\det[EI_N-K_N(x)]$. A simple, transverse, and nondefective one-defect zero crossing at $x=x_{\rm c}$ is characterized locally by
\begin{equation}
P_N(0;x_{\rm c})=0, \quad   \partial_EP_N(0;x_{\rm c})\neq0, \quad  \partial_xP_N(0;x_{\rm c})\neq0,
\label{general_crossing_condition}
\end{equation}
where the first derivative with respect to $E$ ensures that the zero eigenvalue is algebraically simple. Hence, in a neighborhood of $x_{\rm c}$, there is a unique eigenvalue branch $\lambda_{\rm cr}(x)$ satisfying $\lambda_{\rm cr}(x_{\rm c})=0$. The second nonvanishing derivative makes this crossing transverse, since the implicit-function theorem gives
\begin{equation}
\left.
\frac{d\lambda_{\rm cr}}{dx}
\right|_{x_{\rm c}}
=
-
\frac{\partial_xP_N(0;x_{\rm c})}
{\partial_EP_N(0;x_{\rm c})}.
\label{crossing_transversality}
\end{equation}
Now we discuss implications of fermion parity, which is fixed by the defect number. In particular, using $n_j=\nu_j$ on odd sites and $n_j=1-\nu_j$ on even sites gives us
\begin{equation}
\Pi
=
(-1)^{\sum_jn_j}
=
(-1)^{\lfloor N/2\rfloor}
(-1)^{\widehat M}.
\label{parity_defect_relation}
\end{equation}
At a simple zero crossing, the two candidate ground states differ by the occupation of one defect mode and therefore have defect numbers $m$ and $m+1$. Eq. (\ref{parity_defect_relation}) assigns them opposite physical parities. An avoided crossing would require a Hamiltonian term that couples these states, but such a term would have to change their parity. Since $[H,\Pi]=0$, this coupling vanishes, so the two states can consequently have equal energies while remaining independent, even at finite $N$. A perturbation $V$ that breaks defect-number conservation may mix sectors such as $m$ and $m+2$, but if $[V,\Pi]=0$, it cannot mix the crossing states in sectors $m$ and $m+1$. 

\subsection{Alternating finite chain}
\label{subsec:solvable_ground_state_chain}

For a finite chain with $N\geq2$, a solvable realization is obtained by taking $\varepsilon_{2r-1}=a$ and $\varepsilon_{2r}=b$, with $a,b>0$, together with uniform directed pairing amplitudes $\Delta_{j,+}=\Delta_+$ and $\Delta_{j,-}=\Delta_-$, whose product is denoted by $\Gamma=\Delta_+\Delta_-$ (notation not to be confused with $d\Gamma(\cdot)$ used in the exterior-power representation). The diagonal elements of $K_N$ then alternate between $a$ and $-b$, and every off-diagonal product is equal to $\Gamma$. Here it is convenient to introduce the parameters $\delta_\varepsilon=(a-b)/2$, $\overline{\varepsilon}=(a+b)/2$, and $q_\ell=\ell\pi/(N+1)$. As shown in Appendix \ref{appD}, the spectrum consists of the pairs
\begin{equation}
\lambda_{\ell,\pm}
=
\delta_\varepsilon
\pm
\sqrt{
\overline{\varepsilon}^{2}
+
4\Gamma\cos^2q_\ell
},
\quad
\ell=1,\cdots,\left\lfloor\frac{N}{2}\right\rfloor,
\label{finite_chain_spectrum}
\end{equation}
and for odd $N$, these pairs are supplemented by an unpaired level $\lambda_0=a$. Since $a>0$, the latter mode never participates in the ground-state transition. For negative $\Gamma$, the discriminant in Eq. (\ref{finite_chain_spectrum}) first vanishes for $q_1=\pi/(N+1)$. At this point the traceless part of the two-mode block is nonzero and nilpotent (see Appendix \ref{appD}), so the full block is similar to $J_2(\delta_\varepsilon)$ and the degeneracy is an EP. The first EP boundary and the zero-mode crossings are therefore
\begin{eqnarray}
\Gamma_{\rm EP}^{(N)}
&=&
-
\frac{\overline{\varepsilon}^{2}}
{4\cos^2q_1},
\nonumber\\
\Gamma_{{\rm c},\ell}^{(N)}
&=&
-
\frac{ab}{4\cos^2q_\ell},
\quad \quad
\Gamma_{\rm c}^{(N)}
=
\Gamma_{{\rm c},1}^{(N)},
\label{finite_chain_thresholds}
\end{eqnarray}
where $\Gamma_{{\rm c},\ell}^{(N)}$ follows by setting $\lambda_{\ell,+}\lambda_{\ell,-}
=\delta_\varepsilon^2-
(\overline{\varepsilon}^{2}+4\Gamma\cos^2q_\ell)=-ab-4\Gamma\cos^2q_\ell=0$. At this value the two eigenvalues are $0$ and $a-b$, so the zero mode is nondegenerate for $a\neq b$. We focus from now on the case $\ell = 1$, since $\cos^2q_\ell$ decreases with $\ell$ so the first crossing as $\Gamma$ decreases from zero occurs at $\ell=1$. Now for $a\neq b$, Eq. (\ref{finite_chain_thresholds}) implies
\begin{equation}
\Gamma_{\rm EP}^{(N)}
<
\Gamma_{\rm c}^{(N)}
<0,
\label{crossing_before_EP}
\end{equation}
because $\overline{\varepsilon}^{2}-ab=(a-b)^2/4>0$. The zero-mode crossing thus occurs strictly inside the real and diagonalizable domain, so for every finite $N\geq2$ with $a\neq b$, it is quasi-Hermitian, since at $\Gamma=\Gamma_{\rm c}^{(N)}$ one has
\begin{equation}
\overline{\varepsilon}^{2}
+
4\Gamma_{\rm c}^{(N)}\cos^2q_1
=
\frac{(a-b)^2}{4}>0,
\label{crossing_positive_discriminant}
\end{equation}
so the critical $q_1$ subblock has the two distinct eigenvalues $0$ and $a-b$. The crossing is also transverse, since by denoting the zero-crossing branch by $\lambda_{\rm rel}=\lambda_{1,-}$ for $a>b$ and $\lambda_{\rm rel}=\lambda_{1,+}$ for $a<b$, one obtains
\begin{equation}
\left.
\frac{d\lambda_{\rm rel}}{d\Gamma}
\right|_{\Gamma=\Gamma_{\rm c}^{(N)}}
=
\frac{4\cos^2q_1}{b-a}
\neq0.
\label{alternating_crossing_transversality}
\end{equation}
At $\Gamma=\Gamma_{\rm c}^{(N)}<0$, the positivity established for the first mode in Eq. (\ref{crossing_positive_discriminant}) holds for every paired mode: since $\cos^2q_\ell\leq\cos^2q_1$, one has
$\overline{\varepsilon}^{2}+4\Gamma_{\rm c}^{(N)}\cos^2q_\ell\geq(a-b)^2/4>0$.
Thus every pair in Eq. (\ref{finite_chain_spectrum}) has distinct real eigenvalues and the additional level $a$ for odd $N$ is real. The block decomposition (see Appendix \ref{appD} for proof) then shows that $K_N$ is diagonalizable at the crossing and in a neighborhood of it. Its many-body lift in Eq. (\ref{many_body_second_quantization}) has the same properties, so a positive metric exists for the full finite chain. 

\vspace{2mm}

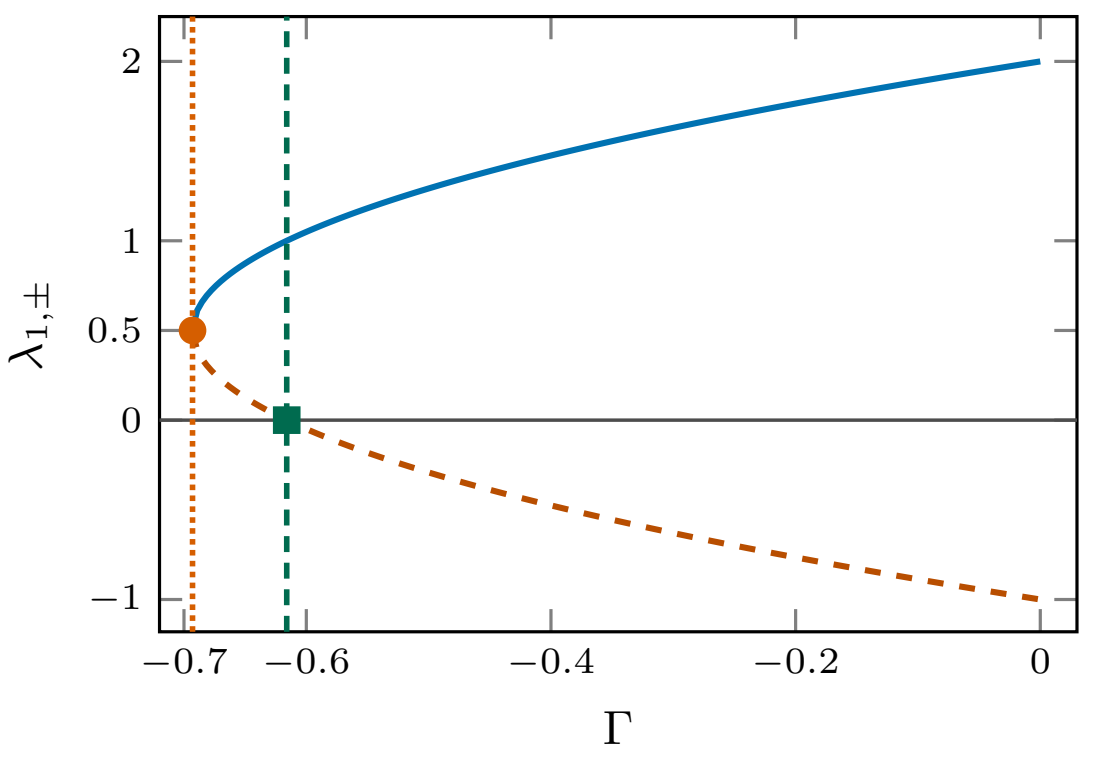
\begin{figure}[t]
\centering
\begin{tikzpicture}

  \definecolor{qhdarkorange}{RGB}{184,78,0}
  \definecolor{qhdarkgreen}{RGB}{0,107,79}

  \pgfmathsetmacro{\GammaEPplot}
    {-2.25/(4*cos(180/7)^2)}
  \pgfmathsetmacro{\GammaCplot}
    {-2/(4*cos(180/7)^2)}

  \begin{axis}[
    width=7.65cm,
    height=5.65cm,
    xmin=-0.72,
    xmax=0.03,
    ymin=-1.18,
    ymax=2.25,
    xlabel={$\Gamma$},
    ylabel={$\lambda_{1,\pm}$},
    xtick={-0.7,-0.6,-0.4,-0.2,0},
    ytick={-1,0,0.5,1,2},
    tick label style={font=\scriptsize},
    label style={font=\small},
    axis line style={line width=0.6pt},
    tick style={line width=0.6pt},
    clip=false
  ]

    \addplot[
      qhblue,
      line width=1.15pt,
      domain=\GammaEPplot:0,
      samples=181
    ]
      {0.5+sqrt(max(
         0,
         2.25+4*x*cos(180/7)^2
       ))};

    \addplot[
      qhdarkorange,
      dashed,
      line width=1.15pt,
      domain=\GammaEPplot:0,
      samples=181
    ]
      {0.5-sqrt(max(
         0,
         2.25+4*x*cos(180/7)^2
       ))};

    \draw[
      black!70,
      line width=0.65pt
    ]
      (axis cs:-0.72,0)--(axis cs:0.03,0);

    \draw[
      qhred,
      densely dotted,
      line width=1.05pt
    ]
      (axis cs:\GammaEPplot,-1.18)--
      (axis cs:\GammaEPplot,2.25);

    \draw[
      qhdarkgreen,
      densely dashed,
      line width=1.05pt
    ]
      (axis cs:\GammaCplot,-1.18)--
      (axis cs:\GammaCplot,2.25);

    \addplot[
      qhred,
      only marks,
      mark=*,
      mark size=2.4pt
    ]
      coordinates {(\GammaEPplot,0.5)};

    \addplot[
      qhdarkgreen,
      only marks,
      mark=square*,
      mark size=2.4pt
    ]
      coordinates {(\GammaCplot,0)};

  \end{axis}
\end{tikzpicture}

\caption{\justifying{Finite-size parity-changing crossing before the EP boundary, shown for $N=6$ with $b=1$ and $a=2$. The solid blue and dashed dark-orange curves represent, respectively, $\lambda_{1,+}=1/2+\sqrt{9/4+4\Gamma\cos^2(\pi/7)}$ and
$\lambda_{1,-}=1/2-\sqrt{9/4+4\Gamma\cos^2(\pi/7)}$.
Starting from $\Gamma=0$ and decreasing $\Gamma$, the occupied lower level reaches zero at
$\Gamma_{\mathrm{c}}^{(6)}=-1/[2\cos^2(\pi/7)]\simeq-0.61596$,
marked by the dark-green dashed line and square. Emptying this level changes the defect number by one, so the two many-body ground states exchanged at the crossing have opposite physical fermion parity. No one-defect eigenvalue coalescence occurs at this crossing, and the two degenerate many-body ground states remain linearly independent. Only at the more negative value $\Gamma_{\rm EP}^{(6)}=-9/[16\cos^2(\pi/7)]\simeq-0.69295$, marked by the red dotted line and circle, do the two one-defect levels
coalesce at $1/2$ and the matrix become defective. Thus the
parity-changing crossing lies strictly inside the real, diagonalizable quasi-Hermitian domain.}}
\label{fig:finite_parity_crossing}
\end{figure}

At $\Gamma=0$, the two branches originate at $-b$ and $a$. If $a>b$, the occupied branch $\lambda_{1,-}$ crosses upward and becomes empty. If $a<b$, the branch $\lambda_{1,+}$ initially empty crosses downward and becomes occupied. In either case, $\widehat{M}$ changes by one and the physical parity reverses. When $a=b$, however, the zero-mode condition coincides with the EP boundary, so we focus strictly on $a \neq b$. The strict separation between the opposite-parity zero crossing and the EP boundary is displayed for a representative finite chain in Fig. \ref{fig:finite_parity_crossing}.

\vspace{2mm}

To pass through the crossing by varying just one real parameter, we pick the parametrization $\Delta_+=z+\gamma$ and $\Delta_-=z-\gamma$ at fixed $\gamma$
\cite{Sinha_2024}. Their product
$\Gamma(z)=\Delta_+\Delta_-=z^2-\gamma^2$ decreases to its minimum $-\gamma^2$ at $z=0$ and then increases, so it can reach the same zero-mode threshold twice. Equating $\Gamma(z)$ to $\Gamma_{\rm c}^{(N)}$
in Eq. (\ref{finite_chain_thresholds}) gives
\begin{equation}
z_\pm=\pm
\left(\gamma^2-
\frac{ab}{4\cos^2q_1}
\right)^{1/2}.
\label{z_crossing_values}
\end{equation}
The two roots are real and distinct when
$\gamma^2>ab/(4\cos^2q_1)$. To move between them while the spectrum remains real and diagonalizable, the minimum $-\gamma^2$
must also stay above the first EP boundary
$\Gamma_{\rm EP}^{(N)}$ in Eq. (\ref{finite_chain_thresholds}).
Together, these conditions give
\begin{equation}
\frac{ab}{4\cos^2q_1}
<
\gamma^2
<
\frac{\overline{\varepsilon}^{2}}
{4\cos^2q_1}.
\label{real_crossing_corridor}
\end{equation}
At the lower equality in Eq. (\ref{real_crossing_corridor}), both the roots merge at $z=0$ and the level only touches zero, so the ground-state defect number does not change. At the upper equality,
$z=0$ reaches the EP boundary and the path between the roots is no longer entirely diagonalizable. Balanced real pairing cannot produce these crossings, since Hermiticity requires $\Delta_+=\Delta_-$ or $\gamma=0$, and then
$\Gamma=z^2\geq0$. Since
$\overline{\varepsilon}^{2}-\delta_\varepsilon^2=ab>0$,
Eq. (\ref{finite_chain_spectrum}) gives
\begin{equation}
\lambda_{\ell,-}<0<\lambda_{\ell,+}
\quad \quad
\text{for all }z\in{\mathbb R}\text{ and all }\ell.
\label{no_Hermitian_crossing}
\end{equation}
Thus, in this alternating-chain family, these crossings require pairing imbalance with $\Delta_+\Delta_-<0$.

\subsection{Spectral Lifshitz criticality at $N\rightarrow\infty$}
\label{criticality_sec}

We now assess the thermodynamic limit $N \rightarrow \infty$. For the alternating chain, the one-defect bands obtained from
Eq. (\ref{finite_chain_spectrum}) become continuous functions of $q\in[0,\pi/2]$, as given by
\begin{equation}
\lambda_\pm(q)
=
\delta_\varepsilon
\pm
\sqrt{
\overline{\varepsilon}^{2}
+
4\Gamma\cos^2q
},
\label{thermodynamic_defect_bands}
\end{equation}
where
$\delta_\varepsilon=(a-b)/2$ and
$\overline{\varepsilon}=(a+b)/2$. Taking
$q_1=\pi/(N+1)\rightarrow0$ in Eq. (\ref{finite_chain_thresholds}) gives
\begin{eqnarray}
&&\Gamma_{\rm c}^{(\infty)}
=
-\frac{ab}{4}, \quad \quad
\Gamma_{\rm EP}^{(\infty)}
=
-\frac{\overline{\varepsilon}^{2}}{4},
\nonumber\\
&&\Gamma_{\rm c}^{(\infty)}
-
\Gamma_{\rm EP}^{(\infty)}
=
\frac{(a-b)^2}{16}>0.
\label{thermodynamic_critical_couplings}
\end{eqnarray}
Thus, even in the thermodynamic limit, the bulk zero-mode transition remains separated from the EP boundary by a finite real-spectrum interval. Consequently, every finite-$N$ approximant remains real and
diagonalizable at the bulk critical coupling $\Gamma_{\rm c}^{(\infty)}$. To parametrize both the sides of the transition, we introduce the signed variable
\begin{equation}
\tau
=
\Gamma_{\rm c}^{(\infty)}-\Gamma.
\label{distance_from_bulk_crossing}
\end{equation}
so that $\Gamma=-ab/4-\tau$. We now lower $\Gamma$ through the bulk threshold
$\Gamma_{\rm c}^{(\infty)}$. On the $\tau>0$ side, we assume $0<\tau<(a-b)^2/16$, keeping
$\Gamma>\Gamma_{\rm EP}^{(\infty)}$ and the bands real and nondefective. Before the threshold ($\tau<0$), Eq. (\ref{thermodynamic_defect_bands}) has
$\lambda_-(q)<0<\lambda_+(q)$ for every $q$: the lower-band modes are occupied and the upper-band modes are empty. At $\tau=0$, one of these bands first reaches zero at $q=0$.
For $\tau>0$, modes near $q=0$ can therefore have the opposite occupation from their pre-transition values. The modes whose occupation changes form an interval $0\leq q<q_F$. We call this interval a ``defect pocket''. In particular, we call it a ``hole'' pocket if previously occupied modes are emptied or a ``particle'' pocket if previously empty modes
are filled. Its boundary $q_F$ is where the changing band has zero energy. Multiplying the two bands in Eq. (\ref{thermodynamic_defect_bands}) gives
$\lambda_+(q)\lambda_-(q)=-ab-4\Gamma\cos^2q$, so setting this product to zero and substituting $\Gamma=-ab/4-\tau$ yields
\begin{equation}
4\Gamma\cos^2q_F=-ab,
\quad \quad
\tan q_F
=
2\sqrt{\frac{\tau}{ab}}.
\label{Fermi_pocket_boundary}
\end{equation}
For $a>b$, $\lambda_-$ becomes positive for $0\leq q<q_F$, so those lower-band modes are emptied and form a hole pocket. For $a<b$, $\lambda_+$ becomes negative there, so those upper-band modes are filled and form a particle pocket. The allowed $q$ values are separated asymptotically by $\pi/N$, so any interval $0\leq q<q_F$ with $q_F>0$ contains a number of modes proportional to $N$. Thus the newly occupied or emptied interval survives in the thermodynamic limit, whereas one finite-size parity crossing changes the occupation of only a single mode.

\vspace{2mm}

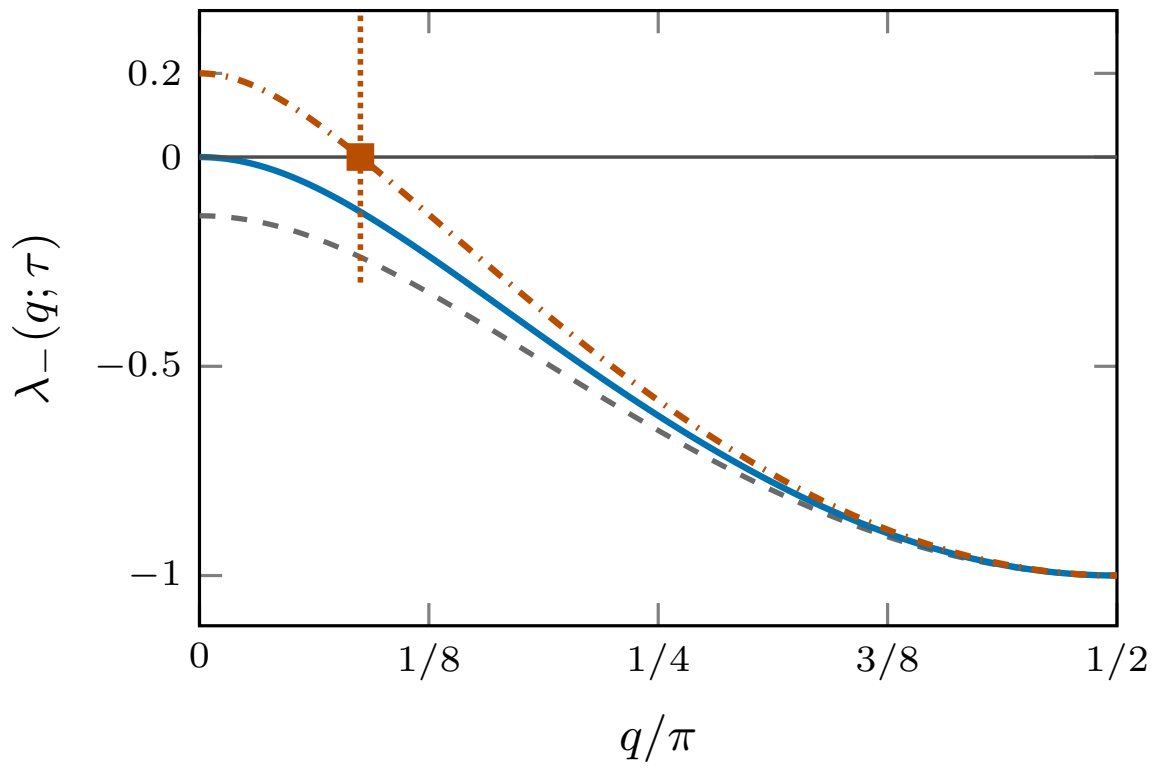
\begin{figure}[t]
\centering
\begin{tikzpicture}

  \definecolor{qhdarkorange}{RGB}{184,78,0}

  \pgfmathsetmacro{\qFplot}
    {atan(2*sqrt(0.04/2))/180}

  \begin{axis}[
    width=7.65cm,
    height=5.65cm,
    xmin=0,
    xmax=0.5,
    ymin=-1.12,
    ymax=0.35,
    xlabel={$q/\pi$},
    ylabel={$\lambda_-(q;\tau)$},
    xtick={0,0.125,0.25,0.375,0.5},
    xticklabels={$0$,$1/8$,$1/4$,$3/8$,$1/2$},
    ytick={-1,-0.5,0,0.2},
    tick label style={font=\scriptsize},
    label style={font=\small},
    axis line style={line width=0.6pt},
    tick style={line width=0.6pt},
    clip=false
  ]

    \addplot[
      qhgray,
      dashed,
      line width=1.0pt,
      domain=0:0.5,
      samples=181
    ]
      {0.5-sqrt(
         0.25
         +2*sin(180*x)^2
         +0.16*cos(180*x)^2
       )};

    \addplot[
      qhblue,
      line width=1.2pt,
      domain=0:0.5,
      samples=181
    ]
      {0.5-sqrt(
         0.25
         +2*sin(180*x)^2
       )};

    \addplot[
      qhdarkorange,
      dash dot,
      line width=1.2pt,
      domain=0:0.5,
      samples=181
    ]
      {0.5-sqrt(max(
         0,
         0.25
         +2*sin(180*x)^2
         -0.16*cos(180*x)^2
       ))};

    \draw[
      black!70,
      line width=0.65pt
    ]
      (axis cs:0,0)--(axis cs:0.5,0);

    \draw[
      qhdarkorange,
      densely dotted,
      line width=1.05pt
    ]
      (axis cs:\qFplot,-0.30)--
      (axis cs:\qFplot,0.35);

    \addplot[
      qhdarkorange,
      only marks,
      mark=square*,
      mark size=2.4pt
    ]
      coordinates {(\qFplot,0)};

  \end{axis}
\end{tikzpicture}

\caption{\justifying{Thermodynamic formation of the Lifshitz hole pocket for $b=1$ and $a=2$. With $\tau=\Gamma_{\mathrm{c}}^{(\infty)}-\Gamma$ and $\Gamma_{\mathrm{c}}^{(\infty)}=-1/2$, the lower band is $\lambda_-(q;\tau)=1/2-\sqrt{1/4+2\sin^2q-4\tau\cos^2q}$, and its negative-energy modes are occupied. The dashed gray, solid blue, and dash-dotted dark-orange curves correspond to $\tau=-0.04$, $\tau=0$, and $\tau=0.04$, respectively. Before the transition, the relevant lower band remains negative and is separated from zero. At the transition it
touches zero at $q=0$, and the positive removal energy obeys $-\lambda_-(q;0)=2q^2+\mathcal{O}(q^4)$, giving $z_{\mathrm{dyn}}=2$. Beyond the transition, the band is positive for $0\leq q<q_F$, so those modes are emptied and form a hole pocket. For the displayed post-transition curve, $q_F=\arctan(2\sqrt{0.04/2})$ and
$q_F/\pi\simeq0.08774$; the dark-orange dotted line and square mark this pocket boundary. The chosen value $\tau=0.04<1/16$ remains on the real, diagonalizable side of the thermodynamic-limit EP boundary.}}
\label{fig:lifshitz_band_transition}
\end{figure}

To describe the energy of changing one mode's occupation, we let $\Xi_{\rm L}(q,\tau)$ be the energy difference relative to its pre-transition occupation. Using $\Gamma=-ab/4-\tau$ from Eqs. (\ref{thermodynamic_critical_couplings}) and
(\ref{distance_from_bulk_crossing}) in Eq. (\ref{thermodynamic_defect_bands}), and noting that $\overline{\varepsilon}^{2}-ab=\delta_\varepsilon^2$, gives the bands as
$\lambda_\pm(q,\tau)=\delta_\varepsilon\pm
\sqrt{\delta_\varepsilon^2+ab\sin^2q-4\tau\cos^2q}$. For $a>b$, emptying an occupied lower-band mode changes the many-body energy by $-\lambda_-$ and for $a<b$, filling an empty upper-band mode changes it by $\lambda_+$. These two energy changes have the same expression, which we denote by
$\Xi_{\rm L}(q,\tau)$:
\begin{eqnarray}
\Xi_{\rm L}(q,\tau)
&=&
\sqrt{
\frac{(a-b)^2}{4}
+
ab\sin^2q
-
4\tau\cos^2q
}
-
\frac{|a-b|}{2}
\nonumber\\
&=&
\frac{abq^2-4\tau}{|a-b|}
+
{\cal O}
\left(q^4,\tau q^2,\tau^2\right).
\label{lifshitz_low_energy_form}
\end{eqnarray}
where in obtaining the second equality, we used $\sin^2q=q^2+{\cal O}(q^4)$ and $\cos^2q=1+{\cal O}(q^2)$, and then expanded the square root about $(a-b)^2/4$ for $a\neq b$. Where $\Xi_{\rm L}<0$, changing the occupation lowers the energy. Its zero gives $q_F\propto\tau^{1/2}$, in agreement
with Eq. (\ref{Fermi_pocket_boundary}). At the transition, we have $\tau=0$ and the energy needed to change a mode's occupation is
$\epsilon(q)=\Xi_{\rm L}(q,0)$. We thus have
\begin{eqnarray}
\epsilon(q)
&=&
\sqrt{
\frac{(a-b)^2}{4}
+
ab\sin^2q
}
-
\frac{|a-b|}{2}
\nonumber\\
&=&
\frac{ab}{|a-b|}q^2
+
{\cal O}(q^4).
\label{quadratic_critical_dispersion}
\end{eqnarray}
showing that the band touches zero quadratically and has the dynamical exponent $z_{\rm dyn}=2$. Varying $\Gamma$ therefore changes the topology of the occupied set of defect modes by nucleating a particle or hole pocket without spontaneous breaking of a local symmetry, identifying this as a one-dimensional spectral Lifshitz transition \cite{Lifshitz_1960}. The evolution of the occupation-changing band through the bulk transition is shown in Fig. \ref{fig:lifshitz_band_transition}.

\vspace{2mm}

There arises in fact a tantalizing bulk thermodynamic effect which we now expose by calculating the ground-state energy per site. Let $E_{\rm gs}(N,\Gamma)$ be the lowest many-body energy of the $N$-site chain, calculated as in Eq. (\ref{general_ground_state_energy}), and let
$e_{\rm gs}(\Gamma) =\lim_{N\rightarrow\infty}E_{\rm gs}(N,\Gamma)/N$. We compare it with $e_{\rm reg}(\Gamma)$, the smooth energy density obtained by keeping the pre-transition occupations
when $\tau>0$, i.e., the lower band remains filled and the upper band remains empty. After the transition, $e_{\rm reg}$ is
only a reference since the actual ground state changes the occupations of modes with $0\leq q<q_F$.

\vspace{2mm}

Using $\Gamma=-ab/4-\tau$ in the bands of
Eq. (\ref{thermodynamic_defect_bands}), we denote their nonnegative square-root term by
\begin{equation}
R(q,\tau)
=
\sqrt{
\delta_\varepsilon^2
+ab\sin^2q
-4\tau\cos^2q
},
\end{equation}
i.e., $\lambda_\pm(q,\tau)=\delta_\varepsilon\pm R(q,\tau)$. As explained in Eq. (\ref{lifshitz_low_energy_form}),
changing a mode's occupation changes its energy by $\Xi_{\rm L}(q,\tau)=R(q,\tau)-|\delta_\varepsilon|$. Inside $0\leq q<q_F$ this quantity is negative, since changing
the occupation lowers the energy by
$|\delta_\varepsilon|-R(q,\tau)$. Since the allowed $q$ values have spacing approximately $\pi/N$, summing this energy reduction over the changed modes and dividing by $N$ gives (see Appendix \ref{appE}) 
\begin{eqnarray}
e_{\rm gs}(\Gamma)-e_{\rm reg}(\Gamma)
&=&
-\frac{1}{\pi}
\int_0^{q_F}
\left[
|\delta_\varepsilon|-R(q,\tau)
\right]dq
\label{ground_state_energy_singularity}
\\
&=&
-\frac{16}
{3\pi|a-b|\sqrt{ab}}\,\tau^{3/2}
+{\cal O}(\tau^{5/2}),
\nonumber
\end{eqnarray}
for $\tau>0$. The power scaling follows from
$q_F=2\sqrt{\tau/(ab)}+{\cal O}(\tau^{3/2})$ in Eq. (\ref{Fermi_pocket_boundary}) and
$|\delta_\varepsilon|-R(q,\tau)=(4\tau-abq^2)/|a-b|+{\cal O}(\tau^2)$ throughout the interval $q=O(\sqrt{\tau})$. For $\tau<0$ no modes change occupation, so $e_{\rm gs}=e_{\rm reg}$ near the transition. Consequently the ground-state energy density and its first derivative with respect to $\Gamma$ are continuous, whereas its second derivative diverges as $\tau^{-1/2}$ from the
$\tau>0$ side, thereby showing a signature of quantum-critical behavior. We conclude by remarking that finite-dimensional quasi-Hermiticity at every finite $N$ does not automatically guarantee a well-defined quasi-Hermitian representation in the infinite-volume limit, so our thermodynamic-limit results are to be understood purely at the spectral level.

\section{Conclusions}\label{disc_sec}

We have established two complementary consequences of an exactly solvable imbalanced Kitaev chain after a staggered particle-hole transformation reduces the many-body problem to fixed-defect exterior powers of a tridiagonal one-defect matrix $K_N$. First, we show that an explicit parameter family reaches a principal size-$N$ Jordan block as the endpoint of a quasi-Hermitian corridor. The induced $m$-defect dynamics has longest Jordan chain $L_m=m(N-m)+1$, giving the maximal many-body order $L_{\rm max}=\lfloor N^2/4\rfloor+1$. Hence many-body Jordan amplification first exceeds the one-defect order at $N=4$. 

\vspace{2mm}

In a separate alternating-chain specialization, a negative-product pairing regime supports simple zero-energy crossings between opposite-parity ground states while the finite chain remains real, diagonalizable, and quasi-Hermitian. For $a\neq b$, the first such crossing occurs strictly before the EP boundary at every finite $N$ and its thermodynamic-limit accumulation point remains separated from the thermodynamic EP boundary by $(a-b)^2/16$. Their thermodynamic-limit accumulation nucleates a defect-particle or defect-hole pocket with $q_F\propto\tau^{1/2}$ and yields a one-dimensional spectral Lifshitz transition with quadratic critical dispersion ($z_{\rm dyn}=2$) and a $\tau^{3/2}$ singular contribution to the ground-state energy density. This provides a negative-product pairing route to spectral Lifshitz criticality that is distinct from both EP coalescence in non-Hermitian Kitaev chains and the supercurrent-driven Lifshitz mechanism known in Hermitian Kitaev superconductors \cite{MedinaCuy_2024}.\\

\noindent
\textbf{Acknowledgements:} A.G. thanks the Department of Physics, IIT Bhubaneswar for hospitality during March-June 2025, during which some related preliminary calculations were performed. We thank Bijan Bagchi for related discussions. The authors have used GPT-Sol 6 for proofreading the manuscript and for verifying the mathematical derivations by the authors. The authors take full responsibility for the validity of the results.

\appendix 

\section{Proof that $K_{N,{\rm EP}}$ has Jordan-block structure of size $N$}\label{appA}

We now prove the size-$N$ Jordan-block assertion associated with the tuning in Eq. (\ref{EPN_parameter_family}). We have $u_j=(K_{N,{\rm EP}})_{j,j+1}$, $v_j=(K_{N,{\rm EP}})_{j+1,j}$, and $u_jv_j=-s^2j(N-j)\neq0$, with $j=1,\cdots,N-1$. Thus every $u_j$ is nonzero, and one may define the invertible diagonal matrix $D={\rm diag}(\alpha_1,\cdots,\alpha_N)$ recursively by
\begin{equation}
\alpha_1=1,
\quad \quad
\frac{\alpha_{j+1}}{\alpha_j}
=
\frac{sj}{u_j}.
\label{EPN_diagonal_similarity}
\end{equation}
The matrix $\widetilde{K}_N=D^{-1}K_{N,{\rm EP}}D$ then has the entries
\begin{eqnarray}
(\widetilde{K}_N)_{jj} &=& s(N+1-2j), \nonumber \\  (\widetilde{K}_N)_{j,j+1} &=& sj, \nonumber \\
(\widetilde{K}_N)_{j+1,j} &=& -s(N-j),
\label{EPN_canonical_matrix}
\end{eqnarray}
where last equality follows directly from
\begin{equation}
(\widetilde{K}_N)_{j+1,j}
=
v_j\frac{\alpha_j}{\alpha_{j+1}}
=
\frac{u_jv_j}{sj}
=
-s(N-j).
\label{EPN_lower_matrix_element}
\end{equation}
To establish the nilpotency of $\widetilde{K}_N$, let us consider the $N$-dimensional space of homogeneous polynomials of degree $N-1$ with basis
\begin{equation}
\varphi_j(x,y)
=
x^{N-j}y^{j-1},
\quad \quad
j=1,\cdots,N.
\label{EPN_polynomial_basis}
\end{equation}
On this space, let us consider the first-order differential operator $\mathcal{X} = (x-y)(\partial_x+\partial_y)$, whose action on $\varphi_j(x,y)$ is 
\begin{equation}
{\cal X}\varphi_j =  (j-1)\varphi_{j-1}
+ (N+1-2j)\varphi_j -
(N-j)\varphi_{j+1},
\label{EPN_polynomial_action}
\end{equation}
where $\varphi_0=\varphi_{N+1}=0$. Thus
$\widetilde{K}_N$ is the matrix representation of $s{\cal X}$. Introducing the invertible linear change of variables
\begin{equation}
u=x-y,
\quad \quad
v=\frac{x+y}{2},
\label{EPN_polynomial_coordinates}
\end{equation}
one obtains
\begin{equation}
\partial_x+\partial_y=\partial_v,
\quad \quad
{\cal X}=u\partial_v.
\end{equation}
Since $u$ is independent of $v$, one has
\begin{equation}
{\cal X}^{r}
=
u^r\partial_v^r
\end{equation}
on the entire polynomial representation space. Every homogeneous polynomial of total degree $N-1$ has degree at most $N-1$ in $v$. Consequently, one has
\begin{equation}
{\cal X}^{N}=0
\end{equation}
as an operator on the full $N$-dimensional representation space. We next consider the vector $v^{N-1}$ which belongs to the same homogeneous polynomial space. For $0\leq r\leq N-1$, one has
\begin{equation}
{\cal X}^{r}v^{N-1}
=
u^r\partial_v^r v^{N-1}
=
\frac{(N-1)!}{(N-1-r)!}
u^rv^{N-1-r}.
\label{EPN_nilpotent_chain}
\end{equation}
In particular, we can write
\begin{equation}
{\cal X}^{N}=0,
\quad \quad
{\cal X}^{N-1}v^{N-1}
=
(N-1)!u^{N-1}\neq0.
\label{EPN_nilpotency_index}
\end{equation}
Therefore the nilpotency index of ${\cal X}$, and hence of $\widetilde{K}_N=s{\cal X}$ for $s\neq0$, is exactly $N$. Since the representation space itself has dimension $N$, there can be only one Jordan block, i.e., 
\begin{equation}
K_{N,{\rm EP}}
\sim
\widetilde{K}_N
\sim
J_N(0).
\label{single_Jordan_block}
\end{equation}
This proves both algebraic multiplicity $N$ and geometric multiplicity one. Finally, one should also note that 
\begin{equation}
{\rm Tr}[K_{N,{\rm EP}}] = s\sum_{j=1}^{N}(N+1-2j) =0,
\label{EPN_reference_energy}
\end{equation}
with $E_\Lambda = -\frac{1}{2}{\rm Tr}[K_{N,{\rm EP}}] =0$, and therefore the coalesced eigenvalue of both $H_{1,{\rm EP}}$ and $H_{N-1,{\rm EP}}$ is zero without any further energy shift.

\section{Quasi-Hermiticity of $H$}\label{appB}

We now show that the one-defect similarity transformation constructed
in Sec. \ref{sec:negative_product_Dyson_map} extends directly to the complete many-body Hamiltonian. For an arbitrary $N\times N$ matrix $A$, we define the corresponding fermionic bilinear
\begin{equation}
\widehat A
=
\sum_{j,k=1}^{N}
A_{jk}d_j^\dagger d_k.
\label{app_fermionic_matrix_lift}
\end{equation}
Using the canonical anticommutation relations in
Eq. (\ref{canonical_anticommutation}), one finds
\begin{equation}
[d_j^\dagger d_k,d_l^\dagger d_m]
=
\delta_{kl}d_j^\dagger d_m
-
\delta_{mj}d_l^\dagger d_k.
\label{app_bilinear_element_commutator}
\end{equation}
Substituting Eq. (\ref{app_bilinear_element_commutator}) into the
commutator of two operators of the form
Eq. (\ref{app_fermionic_matrix_lift}) gives
\begin{equation}
[\widehat A,\widehat B]
=
\widehat{[A,B]}.
\label{app_bilinear_commutator}
\end{equation}
Thus the map $A\mapsto\widehat A$ preserves the matrix commutator.

\vspace{2mm}

We now define
\begin{equation}
\widehat J_\alpha
=
\sum_{j,k=1}^{N}
(J_\alpha)_{jk}d_j^\dagger d_k,
\qquad
\alpha=x,y,z.
\label{app_manybody_spin_generators}
\end{equation}
Eq. (\ref{app_bilinear_commutator}) implies that these many-body operators satisfy the same angular-momentum algebra as the matrices
$J_\alpha$. In particular, one can write
\begin{equation}
[\widehat J_x,\widehat J_z]
=
-i\widehat J_y,
\qquad
[\widehat J_x,i\widehat J_y]
=
-\widehat J_z.
\label{app_manybody_spin_commutators}
\end{equation}
Along the parameter path used in
Sec. \ref{sec:negative_product_Dyson_map}, one has $a_j=s(N+1-2j)$ and therefore $\sum_{j=1}^{N}a_j=0$. So it follows from Eq. (\ref{defect_onsite_energy}) that
$E_\Lambda=0$. Since $K_N(\rho)=2s(J_z+i\rho J_y)$ from Eq. (\ref{negative_product_spin_form}), the number-conserving form in Eq. (\ref{defect_number_conserving_form}) becomes
\begin{equation}
H(\rho)
=
2s\left(
\widehat J_z+i\rho\widehat J_y
\right).
\label{app_full_spin_hamiltonian}
\end{equation}
We now let $\Omega_\rho = e^{\eta\widehat J_x}$ and $\eta=\operatorname{arctanh}\rho$. Because the commutators in Eq. (\ref{app_manybody_spin_commutators})
are identical to those used in Eq. (\ref{spin_hyperbolic_rotation}), the Baker-Campbell-Hausdorff expansion gives
\begin{eqnarray}
e^{\eta\widehat J_x}\widehat J_z
e^{-\eta\widehat J_x}
&=&
\widehat J_z\cosh\eta
-i\widehat J_y\sinh\eta,
\nonumber\\
e^{\eta\widehat J_x}(i\widehat J_y)
e^{-\eta\widehat J_x}
&=&
i\widehat J_y\cosh\eta
-\widehat J_z\sinh\eta.
\label{app_manybody_hyperbolic_rotation}
\end{eqnarray}
Using $\tanh\eta=\rho$ exactly as in
Sec. \ref{sec:negative_product_Dyson_map}, the
$\widehat J_y$ contribution cancels and one obtains the Dyson map
\begin{equation}
\Omega_\rho H(\rho)\Omega_\rho^{-1}
=
2s\sqrt{1-\rho^2} \widehat J_z
\equiv h_\rho.
\label{app_full_Hermitian_similarity}
\end{equation}
Since $J_z$ is Hermitian, Eq. (\ref{app_manybody_spin_generators})
implies $\widehat J_z=\widehat J_z^\dagger$, and hence $h_\rho=h_\rho^\dagger$, so Eq. (\ref{app_full_Hermitian_similarity}) shows that $H(\rho)$ is similar to a Hermitian matrix for $0 \leq \rho < 1$. Furthermore, $J_x=J_x^\dagger$ implies $\widehat J_x=\widehat J_x^\dagger$. Therefore, for
$0\leq\rho<1$, $\eta$ is finite and
$\Omega_\rho=e^{\eta\widehat J_x}$ is Hermitian, strictly positive, and invertible.

\vspace{2mm}

So finally, taking the Hermitian conjugate of
Eq. (\ref{app_full_Hermitian_similarity}) and using $h_\rho^\dagger=h_\rho$ gives quasi-Hermiticity of the many-body Hamiltonian
\begin{equation}
H^\dagger(\rho)\Theta_\rho = \Theta_\rho H(\rho),
\label{app_full_qh_relation}
\end{equation}
where
\begin{equation}
\Theta_\rho
=
\Omega_\rho^\dagger\Omega_\rho
=
e^{2\eta\widehat J_x}>0
\label{app_full_metric}
\end{equation} is also invertible. Hence Eq. (\ref{app_full_qh_relation}) gives $H^\dagger(\rho) = \Theta_\rho H(\rho)\Theta_\rho^{-1}$, proving the quasi-Hermiticity of the complete many-body Hamiltonian for $0\leq\rho<1$.

\section{Exterior powers and differential second quantization}
\label{appC}

For completeness, let us summarize the exterior-power construction used in the many-body analysis \cite{Derezinski_2013}. Let $V=\mathbb{C}^N$ denote the one-defect space with the basis $\{e_1,\cdots,e_N\}$. The $m$-defect sector is naturally identified with the exterior power $\bigwedge^m V$, which has the basis
\begin{equation}
e_{i_1}\wedge\cdots\wedge e_{i_m},
\quad \quad
1\leq i_1<\cdots<i_m\leq N,
\label{exterior_basis_appendix}
\end{equation}
and dimension $\binom{N}{m}$. An invertible one-defect transformation $S\in{\rm GL}(V)$ induces the multiplicative exterior-power transformation $\bigwedge^m S$ according to
\begin{equation}
\left(\bigwedge\nolimits^m S\right)
\left(w_1\wedge\cdots\wedge w_m\right)
=
Sw_1\wedge\cdots\wedge Sw_m.
\label{multiplicative_exterior_power}
\end{equation}
This construction is multiplicative, giving $\bigwedge^m(S_1S_2)=(\bigwedge^mS_1)(\bigwedge^mS_2)$.

\vspace{2mm}

The operator induced by a one-defect generator $K\in{\rm End}(V)$ is instead the differential (or Lie-algebra) representation
\begin{equation}
d\Gamma_m(K)
\equiv
\left.
\frac{d}{dt}
\bigwedge\nolimits^m\!\left(e^{tK}\right)
\right|_{t=0}.
\label{differential_second_quantization}
\end{equation}
Its action on a decomposable exterior vector is
\begin{equation}
d\Gamma_m(K)
\left(w_1\wedge\cdots\wedge w_m\right)
=
\sum_{a=1}^{m}
w_1\wedge\cdots\wedge Kw_a\wedge\cdots\wedge w_m.
\label{induced_exterior_action_appendix}
\end{equation}
Thus $d\Gamma_m(K)$ is not the multiplicative matrix exterior power $\bigwedge^m K$, which would apply $K$ simultaneously to every factor. In particular, differential second quantization preserves commutators instead of matrix products:
\begin{equation}
\left[
d\Gamma_m(K),d\Gamma_m(L)
\right]
=
d\Gamma_m\!\left([K,L]\right).
\label{induced_commutator_appendix}
\end{equation}
The relation to the defect fermions follows from the identification
\begin{equation}
e_{i_1}\wedge\cdots\wedge e_{i_m}
\longleftrightarrow
d_{i_1}^{\dagger}\cdots d_{i_m}^{\dagger}
|\Lambda\rangle .
\label{wedge_fermion_identification}
\end{equation}
Under this identification, the number-preserving quadratic operator
\begin{equation}
d\Gamma(K)
=
\sum_{i,j=1}^{N}
K_{ij}d_i^\dagger d_j
\label{full_differential_second_quantization}
\end{equation}
restricts to $d\Gamma_m(K)$ on the $m$-defect sector. Equivalently, on the complete defect Fock space, one has
\begin{equation}
d\Gamma(K)
=
\bigoplus_{m=0}^{N}d\Gamma_m(K),
\quad \quad
d\Gamma_0(K)=0.
\label{fock_sector_decomposition_appendix}
\end{equation}
The construction also transports one-defect similarities directly to every defect sector. If $K=SJS^{-1}$, then
\begin{equation}
d\Gamma_m(K)
=
\left(\bigwedge\nolimits^m S\right)
d\Gamma_m(J)
\left(\bigwedge\nolimits^m S\right)^{-1}.
\label{induced_similarity_appendix}
\end{equation}
This identity follows by applying $\bigwedge^m$ to $e^{tK}=Se^{tJ}S^{-1}$ and differentiating at $t=0$. Consequently, when $K_{N,{\rm EP}}=S_NJ_N(0)S_N^{-1}$, the Jordan structure in the $m$-defect sector is determined by the induced generator $d\Gamma_m[J_N(0)]$. This is the precise sense in which the principal one-defect Jordan block is lifted to the many-body sectors.

\section{Calculations underlying
Eq. (\ref{finite_chain_spectrum})}
\label{appD}

Let $C={\rm diag}(1,-1,1,-1,\cdots)$ and let $A_N$ denote the adjacency matrix of the open chain. Since all off-diagonal products of $K_N$ are equal to $\Gamma$, for $\Gamma\neq0$ there is an invertible diagonal matrix $D$ that balances the two directed matrix elements on every bond. Let us define the balanced representative
\begin{equation}
\widetilde{K}_N
\equiv
D^{-1}K_ND
=
\delta_\varepsilon I_N
+
\overline{\varepsilon}C
+
\sqrt{\Gamma}A_N.
\label{balanced_alternating_matrix}
\end{equation}
For $\Gamma<0$, the square root may be chosen complex, as only its square enters the spectrum. The case $\Gamma=0$ must be treated
separately because the balancing matrix $D$ need not possess a finite invertible limit there.

\vspace{2mm}

Let $W_N$ denote the off-diagonal part of $K_N$ at $\Gamma=0$, so that
\begin{equation}
K_N(0)
=
\delta_\varepsilon I_N
+
\overline{\varepsilon}C
+
W_N.
\label{alternating_Gamma_zero_decomposition}
\end{equation}
Because the directed amplitudes are uniform and
$\Gamma=\Delta_+\Delta_-=0$, at least one of $\Delta_+$ and $\Delta_-$ vanishes. If $\Delta_-=0$, Eqs. (\ref{KN_offdiagonal}) show that $W_N$ maps the even sublattice into the odd sublattice and annihilates the odd sublattice. If $\Delta_+=0$, it instead maps the odd sublattice into the even sublattice and annihilates the even sublattice. If both the amplitudes vanish, then $W_N=0$. In every case, one can write
\begin{equation}
W_N^2=0,
\end{equation}
and furthermore, every nonzero matrix element of $W_N$ connects opposite sublattices. Since $C$ equals $+1$ on the odd sublattice and $-1$ on the even sublattice, it follows that
\begin{equation}
\{C,W_N\}=CW_N+W_NC=0.
\end{equation}
Using $C^2=I_N$, one therefore obtains the exact operator identity
\begin{eqnarray}
\left[K_N(0)-\delta_\varepsilon I_N\right]^2
&=&
\left(
\overline{\varepsilon}C+W_N
\right)^2
\nonumber\\
&=&
\overline{\varepsilon}^{2}C^2
+
\overline{\varepsilon}\{C,W_N\}
+
W_N^2
\nonumber\\
&=&
\overline{\varepsilon}^{2}I_N.
\label{alternating_Gamma_zero_identity}
\end{eqnarray}
Since $\delta_\varepsilon+\overline{\varepsilon}=a$ and $\delta_\varepsilon-\overline{\varepsilon}=-b$, this is equivalently
\begin{equation}
\left[K_N(0)-aI_N\right]\left[K_N(0)+bI_N\right]
=0.
\label{alternating_Gamma_zero_polynomial}
\end{equation}
The minimal polynomial of $K_N(0)$ therefore divides $(x-a)(x+b)$ and because $a,b>0$, the two roots $a$ and $-b$ are distinct. The minimal polynomial consequently has no repeated root, which proves that $K_N(0)$ is diagonalizable. Its eigenvalue multiplicities can also be read off directly. Ordering the odd-sublattice basis vectors before the even-sublattice basis vectors makes $K_N(0)$ block upper triangular or block lower triangular, depending on which directed amplitude vanishes, with diagonal blocks
\begin{equation}
aI_{\lceil N/2\rceil}
\quad \quad\text{and}\quad \quad
-bI_{\lfloor N/2\rfloor}.
\end{equation}
Thus the eigenvalue $a$ has multiplicity $\lceil N/2\rceil$ and the eigenvalue $-b$ has multiplicity $\lfloor N/2\rfloor$. This proves
Eq. (\ref{finite_chain_spectrum}) at $\Gamma=0$ without appealing to continuity.

\vspace{2mm}

Now the open-chain sine modes $\phi_q(j)=\sin(qj)$ satisfy
\begin{equation}
A_N\phi_q=2(\cos q)\phi_q,
\quad \quad
q=\frac{\ell\pi}{N+1}.
\end{equation}
Moreover, $C^2=I_N$, $\{C,A_N\}=0$, and
$C\phi_q=\phi_{\pi-q}$. For $q\neq\pi/2$, the two-dimensional space
\begin{equation}
{\cal V}_q
=
{\rm span}
\left\{
\phi_q,\phi_{\pi-q}
\right\}
\end{equation}
is therefore invariant under $\widetilde{K}_N$. In the ordered basis $\{\phi_q,\phi_{\pi-q}\}$, its restriction is
\begin{equation}
\left.
\widetilde{K}_N
\right|_{{\cal V}_q}
=
\delta_\varepsilon I_2
+
\left(
\begin{array}{cc}
2\sqrt{\Gamma}\cos q & \overline{\varepsilon}\\
\overline{\varepsilon} & -2\sqrt{\Gamma}\cos q
\end{array}
\right).
\label{alternating_two_mode_block}
\end{equation}
Equivalently, one can write
\begin{equation}
\left[
\left.
\widetilde{K}_N
\right|_{{\cal V}_q}
-
\delta_\varepsilon I_2
\right]^2
=
\left(
\overline{\varepsilon}^{2}
+
4\Gamma\cos^2q
\right)I_2.
\label{alternating_matrix_squared}
\end{equation}
For $\Gamma\neq0$, the two eigenvalues in ${\cal V}_q$ are consequently $\delta_\varepsilon\pm \sqrt{\overline{\varepsilon}^{2}+4\Gamma\cos^2q}$. Because $K_N$ and $\widetilde{K}_N$ are similar when $\Gamma\neq0$, this proves the paired levels in Eq. (\ref{finite_chain_spectrum}) for every nonzero $\Gamma$. Together with Eqs. (\ref{alternating_Gamma_zero_identity}) and
(\ref{alternating_Gamma_zero_polynomial}), it establishes Eq. (\ref{finite_chain_spectrum}) also at $\Gamma=0$. Whenever the discriminant is nonzero, the two eigenvalues of the corresponding two-mode block are distinct and that block is diagonalizable.
When the discriminant vanishes, Eq. (\ref{alternating_matrix_squared}) shows that its traceless part is nilpotent; since $\overline{\varepsilon}>0$, this traceless part is nonzero and hence is similar to $J_2(0)$. The repeated eigenvalue is therefore defective. For odd $N$, the unpaired sine mode has $q=\pi/2$ and support only on the odd sublattice, so $\widetilde{K}_N$ has the eigenvalue $\delta_\varepsilon+\overline{\varepsilon}=a$; by similarity, $K_N$ has the same unpaired eigenvalue.

\section{Calculations underlying Eq. (\ref{ground_state_energy_singularity})}\label{appE}
To obtain the thermodynamic-limit singularity, we use $\delta_\varepsilon=(a-b)/2$ and
$\Gamma=-ab/4-\tau$. The magnitude of the occupation-changing level inside the pocket is
\begin{equation}
|\delta_\varepsilon|-R(q,\tau),
\quad \quad
R(q,\tau)
=
\sqrt{
\delta_\varepsilon^2
+ab\sin^2q
-4\tau\cos^2q
}.
\label{pocket_level_energy}
\end{equation}
The condition
$R(q_F,\tau)=|\delta_\varepsilon|$
gives
$\tan q_F=2\sqrt{\tau/(ab)}$.
Relative to the analytic continuation of the no-pocket state,
the singular contribution to the energy density is therefore
\begin{equation}
e_{\rm gs}-e_{\rm reg}
=
-\frac{1}{\pi}
\int_0^{q_F}
\left[
|\delta_\varepsilon|-R(q,\tau)
\right]dq.
\label{pocket_energy_integral}
\end{equation}
Since $q=O(\sqrt{\tau})$ inside the pocket, one can write
\begin{equation}
|\delta_\varepsilon|-R(q,\tau)
=
\frac{4\tau-abq^2}{|a-b|}
+
{\cal O}(\tau^2),
\label{pocket_energy_expansion}
\end{equation}
while $q_F=2\sqrt{\tau/(ab)}+{\cal O}(\tau^{3/2})$. Substitution into Eq. (\ref{pocket_energy_integral}) gives
\begin{eqnarray}
e_{\rm gs}-e_{\rm reg}
&=&
-\frac{1}{\pi|a-b|}
\int_0^{2\sqrt{\tau/(ab)}}
\left(
4\tau-abq^2
\right)dq
+
{\cal O}(\tau^{5/2})
\nonumber\\
&=&
-
\frac{16}
{3\pi|a-b|\sqrt{ab}} \tau^{3/2}
+
{\cal O}(\tau^{5/2}),
\end{eqnarray}
which establishes Eq.~(\ref{ground_state_energy_singularity}).

\end{document}